\documentclass[journal=jpcl,manuscript=article]{achemso}

\usepackage[version=3]{mhchem} 
\usepackage[T1]{fontenc}       
\usepackage{multirow}
\usepackage{color}


\author{Erxun Han}
\altaffiliation{Contributed equally to this work}
\affiliation{School of Physics, Peking University, Beijing 100871, China}
\alsoaffiliation
{Interdisciplinary Institute of Light-Element Quantum Materials and Research Center for Light-Element Advanced Materials, Peking University, Beijing 100871, People's Republic of China}

\author{Wei Fang}
\altaffiliation{Contributed equally to this work}
\affiliation{State Key Laboratory of Molecular Reaction Dynamics and Center for Theoretical Computational Chemistry, Dalian Institute of Chemical Physics, Chinese Academy of Sciences, Dalian 116023, P. R. China.}
\alsoaffiliation{Department of Chemistry, Fudan University, Shanghai 200438, China}
\alsoaffiliation{Laboratory of Physical Chemistry, ETH Zurich, CH-8093 Zurich, Switzerland}

\author{Michail Stamatakis}
\affiliation{Thomas Young Center and Department of Chemical Engineering, University College London, Torrington Place, London WC1E 7JE, United Kingdom}

\author{Jeremy O. Richardson}
\affiliation{Laboratory of Physical Chemistry, ETH Zurich, CH-8093 Zurich, Switzerland}

\author{Ji Chen}
\email{ji.chen@pku.edu.cn}
\affiliation{School of Physics, Peking University, Beijing 100871, China}
\alsoaffiliation
{Interdisciplinary Institute of Light-Element Quantum Materials and Research Center for Light-Element Advanced Materials, Peking University, Beijing 100871, People's Republic of China}
\alsoaffiliation{Frontiers
Science Center for Nano-Optoelectronics, Peking University, Beijing 100871, People's Republic of China}
\alsoaffiliation
{Collaborative Innovation Center of Quantum Matter, Beijing 100871, People’s Republic of China}

\title{
Quantum Tunnelling Driven H$_2$ Formation on Graphene
}

\begin{document}

\begin{tocentry}

\centering
\includegraphics[width=8.3cm]{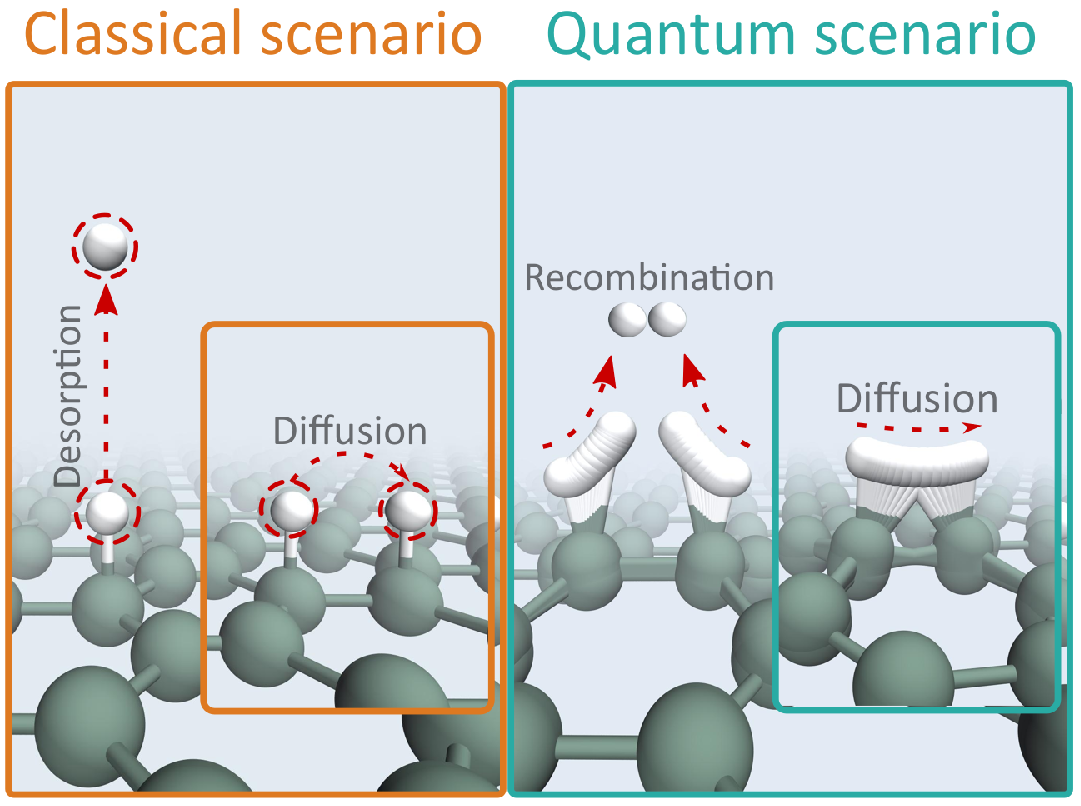}

\end{tocentry}

\clearpage
\begin{abstract}

It is commonly believed that it is unfavourable for adsorbed H atoms on carbonaceous surfaces to form H$_2$ without the help of incident H atoms.
Using ring-polymer instanton theory to describe multidimensional tunnelling effects, combined with \textit{ab initio} electronic structure calculations, we find that these quantum-mechanical simulations reveal a qualitatively different picture.
Recombination of adsorbed
H atoms, which was believed to be irrelevant at low temperature due to high barriers, is enabled by deep tunnelling, with reaction rates enhanced by tens of orders of magnitude.
Furthermore, we identify a new path for H recombination that 
proceeds via multidimensional tunnelling, but would have been predicted to be unfeasible by a simple one-dimensional description of the reaction.
The results suggest that hydrogen molecule formation at low temperatures are rather fast processes that should not be ignored in experimental settings and natural environments with graphene, graphite and other planar carbon segments.
\end{abstract}


\clearpage

Quantum tunnelling, the classically-forbidden transmission of particles through high-energy barriers, plays an important role in many 
physical, chemical and biological processes
\cite{pu_multidimensional_2006,jonsson_simulation_2011,meisner_atom_2016, richardson_concerted_2016}.
%
%
%
Quantum tunnelling can induce phase transitions \cite{fang_quantum_2019}, enhance proton transfer rates\cite{drechsel-grau_quantum_2014,litman_elucidating_2019,lamTheoryElectrochemicalProtonCoupled2020,sakaushiQuantumProtonTunneling2019,sakaushiQuantumtoClassicalTransitionProton2018}, and change the dynamics of water on surfaces \cite{fang_origins_2020}, to name just a few.
Despite its importance, the vast majority of previous studies neglect quantum tunnelling effects, in part due to the challenges involved in accurate computational modeling of multidimensional tunnelling.
While such an approximation is valid in many circumstances, particularly for 
processes at high temperatures, it can be problematic at low temperatures and especially for processes involving hydrogen (see \textit{e.g.}\ refs. \cite{fang_quantum_2019,sims_gas-phase_1995}).

Among all the interesting processes involving hydrogen, adsorption and H$_2$ formation on graphene is of great importance, and has garnered enormous interest and enthusiasm in many fields.
%
%
For instance, graphene/graphite based materials can be used as a promising atomic hydrogen storage ``warehouse''
\cite{elias_control_2009}.
%
The abundance of H$_2$ in outer space is heavily linked to H$_2$ formation on carbonaceous surfaces of interstellar dust grains
\cite{vidali_h-2_2013}.
%
More recently, H$_2$ formation on graphene has also been attributed to enabling the unexpected permeability of H$_2$ through graphene \cite{sun_limits_2020}.
Therefore, the interaction of hydrogen with graphene/graphite, polycyclic aromatic hydrocarbons (PAHs), and other carbonaceous surfaces has been the focus of a large number of studies, see e.g. refs
\cite{morisset_quantum_2004,hornekaer_metastable_2006,hornekaer_clustering_2006,casolo_insights_2013,davidson_cooperative_2014,goumans_hydrogen-atom_2010,petucci_formation_2018} for several notable examples.
%

Hydrogen atoms can adsorb on graphene/graphite surface in a stable chemisorbed state and a less stable physisorbed state \cite{davidson_cooperative_2014}.
Previous studies have established a picture of H$_2$ formation on graphene/graphite
\cite{vidali_h-2_2013}, 
consisting of two mechanisms: (i) collision of an incident H atom with a chemisorbed H atom on the surface, known as the Eley--Rideal (ER) mechanism, and (ii) combination of two physisorbed H atoms, which is a Langmuir--Hinshelwood (LH) type mechanism.
It has been long believed that the LH type mechanism for two chemisorbed H atoms is almost impossible in the cold (and even the moderately warm) regions of interstellar medium, where the temperature ranges from \textit{ca.}\ 10--200 K\@.
However, this established picture is mostly based on classical-mechanical arguments that the potential-energy barriers for the diffusion and recombination of chemisorbed H atoms are too high to be relevant \cite{wakelam_h-2_2017}.
Meanwhile, experimental measurements have indicated that the rate of hydrogen molecule formation is much larger than predictions from simple models based on classical theory \cite{pirronello_laboratory_1997}.
It remains to be unveiled whether quantum mechanics will rewrite our understanding of the mechanism of H$_2$ formation on graphene. 

The past decade saw a blossom of advances in methodology for treating nuclear quantum effects (NQEs), in particular the development of path integral based methods \cite{markland_nuclear_2018,richardson_perspective_2018}.
This contributed to a significant leap in the predictive power of computational modelling and vastly enriched our understanding of NQEs in a wide range of systems \cite{ceriotti_nuclear_2016}. 
%
%
 Using these methods the adsorption of single H atom on PAH and graphene 
has been investigated, and exciting results were reported on how NQEs significantly accelerate the hydrogen adsorption process at low temperatures \cite{goumans_hydrogen-atom_2010,davidson_cooperative_2014,jiangImagingCovalentBond2019a}.
%
These encouraging findings suggest that there is still a lot to be learned regarding the quantum nature of H$_2$ formation on graphene/graphite.
However, a full quantum mechanical treatment of hydrogen molecule formation on surface is a much more challenging task than the study of the adsorption of a single hydrogen atom, demanding a multidimensional tunnelling treatment in conjunction with an on-the-fly \textit{ab initio} treatment of many surface atoms. 
%
%
Although path integral molecular dynamics based methods are well implemented within an \textit{ab initio} framework, they remain computationally very expensive at low temperatures.
Instanton theory however provides a viable alternative which is well suited to low-temperature calculations in the condensed phase\cite{richardson_ring-polymer_2018}. 

In this work, we report a thorough quantum-mechanical investigation of H$_2$ formation on graphene, with the aim of understanding NQEs in H$_2$ formation on graphene and other carbonaceous surfaces.
We employ ring-polymer instanton theory
\cite{richardson_ring-polymer_2018,richardson_ring-polymer_2009,anderssonComparisonQuantumDynamics2009a,kaestnerTheorySimulationAtom2014a},
a state-of-the-art method for modeling multidimensional quantum tunnelling in chemical reactions that can capture the mechanistic difference between quantum and classical reaction pathways, in combination with \textit{ab initio} electronic structure calculations performed on-the-fly.
Overall, we see that the H$_2$ formation rates can be increased by tens of orders of magnitude due to quantum tunnelling, allowing the classically forbidden LH mechanism at chemisorption sites. 
It is of particular interest that multidimensional tunnelling effects, such as ``quantum corner-cutting'', play a crucial role.
%
%
Finally, we employ kinetic Monte Carlo (KMC) simulations to establish a complete picture of the H$_2$ formation processes on graphene, taking into account all competing processes.
The new H$_2$ formation process identified improves significantly the comparison of theory to available experimental data; hence, we propose possible experimental settings to further validate our model.

Our density-functional theory calculations were carried out using the Vienna \textit{ab initio} simulation package (VASP) \cite{kresse_efficient_1996,kresse_efficiency_1996}.
The Perdew-Burke-Ernzerhof (PBE) \cite{perdew_generalized_1997} exchange-correlation functional was used along with the D3 correction \cite{grimme_consistent_2010} to account for van der Waals interactions. 
The climbing image nudged elastic band (CI-NEB) method \cite{henkelman_climbing_2000} was used to obtain the potential energy barriers and minimum energy pathways (MEP) in mass-weighted coordinates. 
Quantum tunnelling rates and pathways were computed with ring-polymer instanton rate theory \cite{richardson_ring-polymer_2009,richardson_ring-polymer_2018} and extrapolated to the infinite-bead limit \cite{beyer_quantum_2016}. 
The multidimensional instantons were obtained via first-order saddle-point optimisations 
at different temperatures with the total force converged to below 0.02 eV$\cdot$\AA$^{-1}$. 
The potential energy surface was calculated on-the-fly with DFT, performed using a python wrapper. 
A total of 64 beads were used to represent the instantons, which was found to give converged results. 
We used the GT-KMC framework as implemented in the software Zacros \cite{stamatakis_graph-theoretical_2011,nielsen_parallel_2013,ravipati_caching_2020} to simulate hydrogen formation.
More computational details and convergence tests are presented in supplementary information (SI). 



\begin{figure}[htbp]
    \centering
    \includegraphics[scale=0.07]{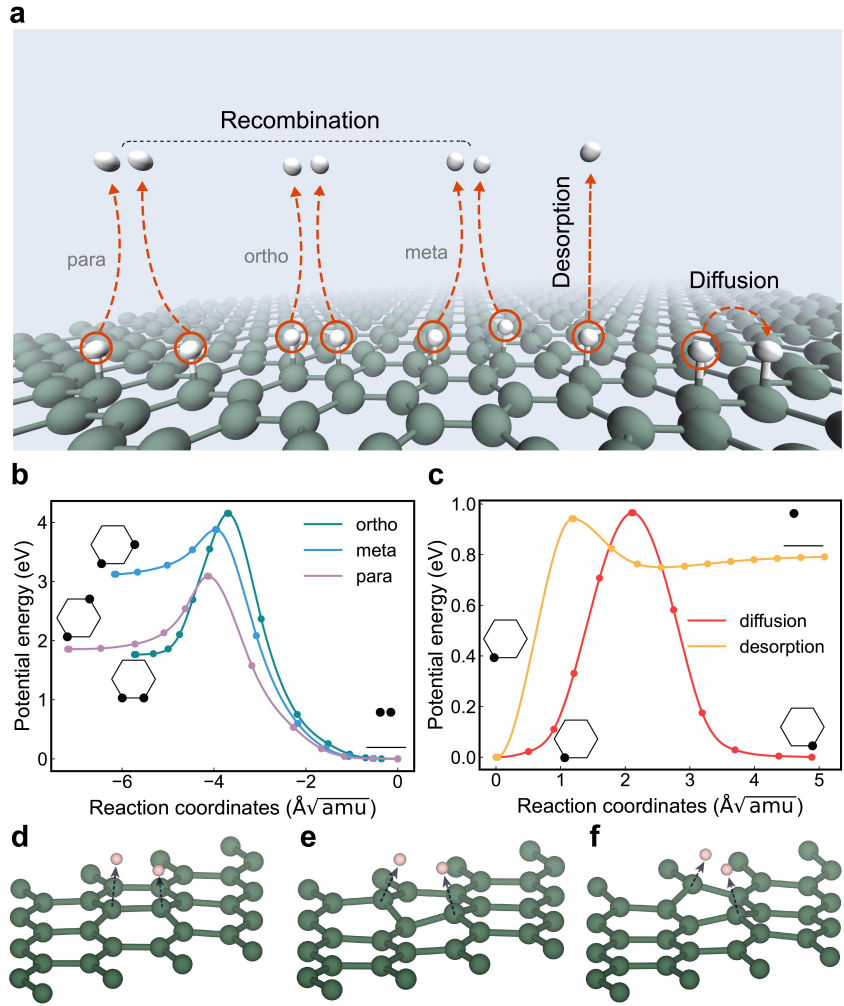}
        \caption{\textbf{
        Classical pathways of key processes of hydrogen formation on graphene.} 
        (a) A schematic showing the Langmuir-Hinshelwood recombination, diffusion, and desorption processes on graphene. 
        Hydrogen and carbon atoms are coloured white and gray-green, respectively.
        Potential energy profile along the (mass-weighted) intrinsic reaction coordinate \cite{fukui_path_1981} for (b) H recombination and (c) H diffusion and desorption/adsorption on graphene.
        The insets illustrate the H positions at the corresponding points on the potential energy profiles.
        (d-f) Geometries of the transition states of H recombination processes starting from ortho, meta and para adsorption sites. 
        }
    \label{figure1}
\end{figure}

An overview of the system and the processes that can occur on the surface is presented in Figure~\ref{figure1}a. 
As a single H atom approaches a graphene surface (or similarly on graphite), it first forms a physisorbed state under the effect of van der Waals (vdW) interactions, with an energy well of 40 meV at the height of $\sim$3.0 \AA~above the surface \cite{davidson_cooperative_2014}. 
Then as the H atom moves closer to the surface, at the height of $\sim$1.5 \AA, it finds itself trapped in a chemisorbed state, which is $\sim$0.75 eV more stable than the physisorbed state (Figure~\ref{figure1}c). 
The C-H bond length is 1.13 \AA, with the carbon atom puckering out of the graphene plane by 0.35 \AA, featuring a sp$^3$ hybridisation \cite{verges_trapping_2010}.
A relatively low barrier of 0.2 eV exists for the transition from physisorption to chemisorption, although it was believed to be large enough that this transition is very difficult 
below room temperature \cite{sha_first-principles_2002}.
However, it has been shown that there are many factors, including nuclear quantum effects, that can dramatically increase the likelihood of H atoms chemisorbing on graphene \cite{davidson_cooperative_2014,goumans_hydrogen-atom_2010,hornekaer_clustering_2006}.
For chemically adsorbed H dimers on graphene, according to the adsorption sites of two H atoms on a hexagon, there are three possible configurations, labeled as ortho, meta, and para as shown in Figure~\ref{figure1}a. 
The most stable adsorption site is the ortho site, which has an adsorption energy of $-$2.83 eV (defined as $E_\text{ad}^{(n)}=E_\text{tot}^{(n)}-n\times E_\text{H}-E_\text{gra}$, where $E_\text{tot}^{(n)}$ is the total energy of the system with $n$ adsorbed H atoms, $E_\text{H}$ is the energy of a free H atom, and $E_\text{gra}$ is the energy of the relaxed graphene surface).
The para site is slightly less stable (by 0.09 eV), while the meta site is the least stable (1.36 eV less stable than the ortho site).
The computed formation energy of H$_2$ is $-4.53$ eV and thus, 
as shown in Figure ~\ref{figure1}b, recombination into a H$_2$ molecule is an exoergic process in each case.
%

\begin{figure}[htbp]
    \centering
    \includegraphics[scale=0.075]{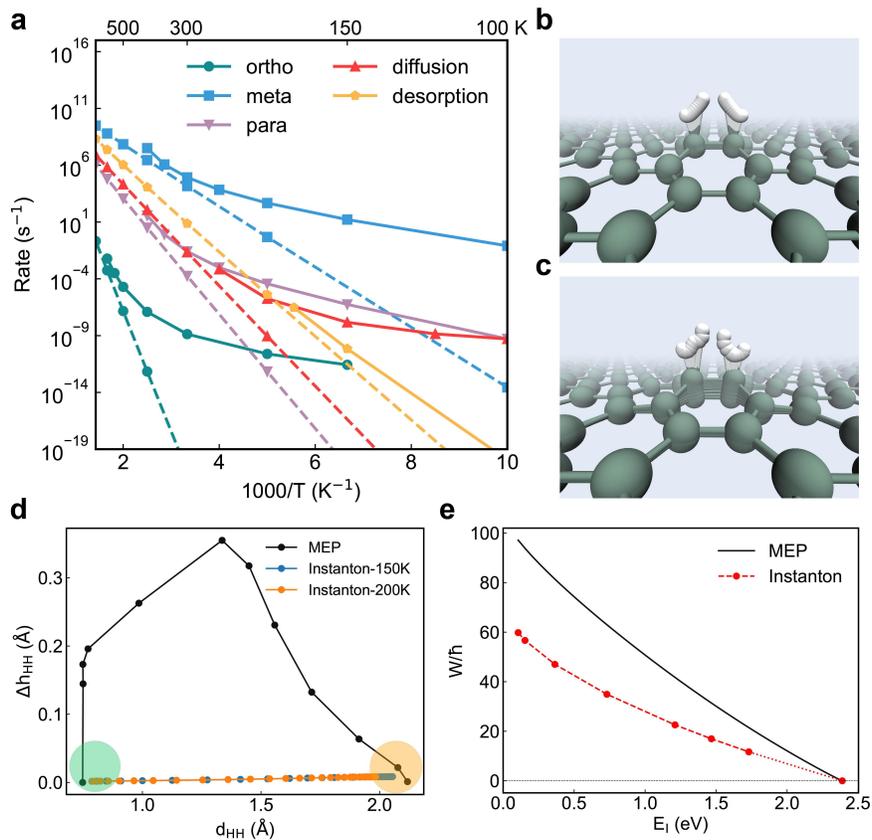}
    \caption{\textbf{
    Quantum and classical reaction rates and pathways.} 
    (a) On-the-fly instanton rates (solid lines) and Eyring transition-state theory (TST) rates (dashed lines) for H recombination, diffusion, and desorption.
    Geometry of (b) the instanton trajectory and (c) the minimum energy pathway (MEP) for H recombination from the ortho site at 150 K\@. 
    (d) Comparison of the MEP and the instanton pathways within a 2D representation.
    d$_{\rm HH}$ represents the distance between the two H atoms and $\Delta$h$_{\rm HH}$ is the height difference of the atoms with respect to the graphene plane.
    The green region indicates the final state and the orange region indicates the initial state.
    (e) Comparison of the abbreviated action ($W$) (see SI eq 2) calculated along the minimum energy pathway (MEP) and along instanton trajectories.
    $E_{\rm I}$ represents the instanton energy or the energy along the reaction coordinate.
    }
    \label{figure2}
\end{figure}

%
We first reexamine the classical picture of H$_2$ formation from chemisorption sites. 
All three H dimers can in principle recombine to form H$_2$ molecules; however, due to the existence of high barriers, it has been long believed that such a process is almost impossible at low temperatures \cite{hornekaer_clustering_2006}.
We identified the minimum energy pathway (MEP) for the three H dimer recombination processes, using the climbing image nudged elastic band (CI-NEB) method.
The potential energy profiles along the MEP are shown in Figure~\ref{figure1}b, and the geometries of the classical transition states are presented in Figure~\ref{figure1}d-f.
The barriers computed in this work agree well with previous studies \cite{borodin_hydrogen_2011,hornekaer_clustering_2006}. 
Among all the H$_2$ formation processes, the one starting from the meta configuration has the lowest energy barrier (0.756 eV).
However, since the meta site is energetically less stable than the other sites, at thermal equilibrium, one might expect to find only a tiny fraction of H dimers at the meta configuration, at least in the limit where H atoms on the surface are sparse, which would make the effect of the meta pathway negligible.
However, if the meta dimer is stabilized, \textit{e.g.}\ , via the formation of stable H atom trimers that contain the meta configuration, then the meta pathway may become an important contributor to the overall process.
Classical mechanics also predicts that the ortho path is unfavourable, as it has the largest potential energy barrier of 2.389 eV\@.
Therefore, one can conclude that, classically, H$_2$ recombination mostly likely happens via the para path, with an energy barrier of 1.234 eV\@.
The classical reaction rate, estimated using Eyring transition-state theory (TST) \cite{rooney_eyring_1995}, is $\sim10^{-4}$ s$^{-1}$ for H recombination from the para site at room temperature and rapidly decreases to $\sim10^{-13}$ s$^{-1}$ at 200 K (Figure~\ref{figure2}a).
This means that classically, recombination processes are unlikely to play a significant role in low-temperature environments.

To shed light on the complete process of H$_2$ formation, we also computed atomic hydrogen diffusion and desorption MEPs (Figure~\ref{figure1}c).
The potential energy barriers of the two processes are almost identical (0.966 eV for diffusion and 0.942 eV for desorption), meaning that classically, desorption and diffusion events occur with similar frequencies.
%
This suggests that long-range diffusion of chemisorbed H atoms on graphene is almost impossible classically \cite{jeloaica_dft_1999}, which would imply that H clusters on graphene cannot be formed via surface diffusion.
Nevertheless, it has been shown that H clusters can form directly on graphene or graphite due to preferential sticking into specific adsorbate structures \cite{hornekaer_clustering_2006}.
The desorption process also has a lower barrier than the H$_2$ formation processes from the stable para and ortho sites, suggesting that recombination is much less likely than for a single H atom to desorb from the surface.
These findings fully agree with the previous studies based on classical mechanics indicating that the LH mechanism is gravely unfavourable for chemisorbed H atoms \cite{wakelam_h-2_2017}.


Due to the light mass of hydrogen, the adsorption, diffusion, and recombination processes are quantum mechanical in nature.  
For this reason we must account for nuclear quantum effects, particularly tunnelling, in our simulations, especially at low temperatures.
As a first step towards understanding the importance of tunnelling, we computed the crossover temperature \cite{gillan_quantum_1987,mills_generalized_1997}, given by $T_{\rm c}\approx\frac{\hbar\omega^\ddagger}{2\pi k_{\rm B}}$, where $\omega^\ddagger$ is the magnitude of the imaginary frequency at the transition state.
$T_{\rm c}$ is the temperature at which the instanton (which represents an optimal tunnelling pathway) starts to delocalize across the barrier and thus it provides an estimate of the temperature below which tunnelling should be considered in a given process.
The three H recombination pathways all have high values of $T_{\rm c}$ above 400 K (see SI Section II), suggesting that H$_2$ formation will exhibit strong tunnelling effects even at room temperature.
The temperatures in interstellar environments and many experimental settings are well below the crossover temperature, such that H$_2$ formation is driven by very deep tunnelling, which would fundamentally invalidate the classical picture.

Using ring-polymer instanton theory with the potential energy surface generated on-the-fly with DFT, we thoroughly investigated the role of tunnelling in this system over a wide range of temperatures to uncover the quantum picture.
The technical details necessary to overcome computational challenges in these calculations and the validation tests are discussed in SI Section I.
For the diffusion and desorption processes, quantum effects flipped the table.  
Unlike in the classical picture, the instanton rate for the diffusion process greatly exceeds that of the desorption process at temperatures below 200 K (see Figure~\ref{figure2}a).
In fact, the desorption process shows only minor tunnelling contributions and the instanton rate \bibnote{Note that the true desorption rate will be even slower than that predicted by instanton theory.  This is because instanton theory only models the transition to the physisorbed state and H atoms that reach the physisorbed state via tunnelling do not necessarily have the energy to escape the physisorption well.} displays a near Arrhenius behaviour over all temperatures considered, while the diffusion rate is strongly non-Arrhenius and the rate becomes weakly temperature dependent at $\sim$100 K, which is a signature of deep tunnelling.
Therefore, in the quantum picture, tunnelling can enable long-range diffusion for a H atom, promoting its mobility even at low temperatures, without being desorbed from the surface.
Quantum diffusion can therefore also facilitate the formation of H clusters on graphene.
The inability of tunnelling to accelerate the desorption process is because it is endothermic by 0.75 eV\@.
Even though tunnelling can accelerate the barrier crossing, it cannot break energy conservation, meaning that the reactant must still be thermally activated by at least 0.75 eV before it can desorb from the surface.
Tunnelling also dominates the H recombination processes.
Instanton rates show strong non-Arrhenius behaviour for all three recombination pathways (Figure~\ref{figure2}a).
For the ortho pathway, even at room temperature, tunnelling plays a significant role in increasing the rate by $\sim$10 orders of magnitude.
The extraordinarily high tunnelling factor results from a combination of features in this pathway, such as short path length, high barrier, high $T_\text{c}$, and strong corner-cutting effects, which all act to enhance the tunnelling factor.
At 150 K, the tunnelling factor for this pathway increases by 44 orders of magnitude.
Strong tunnelling is also seen in the para pathway at low temperature, i.e. at 100 K where tunnelling increases the rate by almost 30 orders of magnitude.
However, compared to the ortho pathway, the increase is less dramatic mainly due to the longer tunnelling distance.
As a result, the rate difference between the two pathways (ortho and para) is greatly reduced in the quantum scenario.
This means that the para pathway may no longer be the only favorable recombination pathway and that the ortho pathway could also play an important role in the quantum scenario, especially bearing in mind that it will be more populated at thermal equilibrium.
There is also a significant tunnelling effect for the meta pathway, albeit to a lesser extent than the other two pathways.
Multidimensional tunnelling effects, namely, corner-cutting  \cite{marcus_new_1977,chapman_semiclassical_1975} where the quantum tunnelling pathway deviates from the classical MEP, also play a crucial role in the H recombination processes.
The classical reaction pathway for the ortho site (Figure~\ref{figure2}c) involves a stepwise mechanism where one H atom first breaks the chemical bond to the graphene surface, then recombines with the other H atom to form H$_2$, which leaves the surface.
This mechanism features an asymmetric transition state, where one H is further away from the surface than the other H atom.
Yet the mechanism qualitatively changes in the quantum scenario, featuring a symmetric tunnelling pathway where the two H atoms tunnel in a concerted fashion, simultaneously approaching each other and leaving the graphene surface (Figure~\ref{figure2}b).
To clearly present the mechanistic change, we show a projection of the pathways onto two generalised coordinates in Figure~\ref{figure2}d.  It can be seen that the height difference between the two H atoms varies along the classical pathway, but the quantum pathway remains symmetric (within the level of numerical convergence).
The qualitative difference between the classical and quantum pathways would invalidate all the MEP based tunnelling correction methods, which are commonly used to predict tunnelling rates in chemical reactions \cite{skodje_general_1981,greene_investigation_2016}.
In fact, the abbreviated action, which characterises the difficulty of transmission through a tunnelling pathway, is significantly lower on the optimal tunnelling pathways (instantons) than on the MEP (Figure~\ref{figure2}e).
This indicates that MEP based methods could underestimate the tunnelling rate by $\sim$10 orders of magnitude for the ortho pathway in the deep tunnelling case (see SI Section III).
The ortho pathway is not the only pathway impacted by strong quantum corner-cutting effects.
We show in SI Section II that the instanton path deviates from the MEP for all three H recombination pathways.
For the para pathway, despite no qualitative mechanistic difference between the classical and the quantum scenario, using the MEP still severely underestimates tunnelling compared to the instanton theory.
By carefully comparing the MEP and the instantons, we find that the quantum corner-cutting effects in the para pathways mainly come from the fact that the optimal tunnelling pathway makes a compromise between minimising the potential energy along the path and trying to avoid tunnelling in the heavier C atoms, which is unfavourable. 
Indeed, we can see that even for the deep tunnelling instantons at the lowest temperature studied, only limited carbon tunnelling is observed (see Figure~\ref{figure2}b).
Overall, the strong quantum corner-cutting effects are widely observed in H atom processes on graphene with quantum tunnelling considered, which reveal the qualitative difference between the quantum and classical reaction pathways in multidimensional space.
%


Moreover, the dominance of nuclear quantum tunnelling predicted in our study is expected to lead to strong kinetic isotope effects (KIEs) \cite{carpenter_kinetic_2010,paris_kinetic_2013}.  
Due to the fact that the richness of deuterium (D) is low (less than 1$\%$), the most likely process would be the formation of HD, while D$_2$ formation would be very rare.
Therefore, we analyse the formation of HD molecules via a Langmuir-Hinshelwood (LH) type mechanism from chemisorbed hydrogen and deuterium using multidimensional instanton theory.
The instanton trajectories for HD formation becomes ``asymmetric'' (see Figure S6), as the tunnelling distance of H and D atoms differ due to their mass differences (D is heavier and less likely to tunnel).
Despite this, the HD tunnelling pathway resembles the H$_2$ tunnelling pathway rather than the MEP, even for the ortho pathway, suggesting that the corner-cutting effects also play a significant role in HD formation.
%
%
At 150 K the rates of HD formation via ortho and para pathways decrease by 3 orders of magnitude compared to the formation rates of H$_2$ (Table 1).
However, the gap between ortho and para rates widens slightly further for HD formation, suggesting that the ortho path is less important compared to H$_2$ formation.
We also expect that recombination rates are further decreased when we consider two deuterium (or tritium) atoms forming D$_2$ (or T$_2$).
Based on the Bell-Limbach tunnelling model \cite{limbach_arrhenius_2006}, the formation rate of D$_2$ would be more than 6 orders of magnitude lower than the formation rates of H$_2$.
%
%
These results are in line with the current understanding that in outer space the isotopic fractionation of deuterium goes through a different path involving other larger molecules \cite{watson_isotope_1977}.
The strong isotope effects also suggest that delicate isotopic control of H$_2$ formation may be achievable.

\begin{table}
\caption{Multidimensional Instanton Rates for H$_2$ and HD Formation Processes at 100 K (150 K for Formation on the Ortho Site).\\}
\begin{tabular}{l|cc}
\hline\hline
\multirow{2}{*}{Process} & \multicolumn{2}{c}{Multidimensional instanton rates (s$^{-1}$)} \\ \cline{2-3} 
 & H$_2$ & HD \\ \hline
Ortho & $3.25\times10^{-12}$ (150 K) & ~~~~$8.97\times10^{-16}$ (150 K)\\
Meta & $1.03\times10^{-1}$ & $1.54\times10^{-3}$ \\
Para & $6.82\times10^{-10}$ & $8.28\times10^{-13}$ \\ \hline\hline
\end{tabular}
\label{tab3}
\end{table}


The previous sections discussed the H$_2$ formation rates from H dimers, which corresponds to the situation in the sparse-coverage limit.
However, H-rich regions may also exist, and
since H$_2$ formation on graphene involves several sub-processes, to better understand the overall mechanism when the H atoms are not sparse, we further employ Graph-Theoretical Kinetic Monte Carlo (GT-KMC) simulations \cite{stamatakis_graph-theoretical_2011,nielsen_parallel_2013,ravipati_caching_2020} to calculate the occurrence frequencies for the different sub-processes.
The GT-KMC simulations are initialized with H atoms randomly adsorbed on the lattice
and a queue containing all the possible processes for the given configurations. 
The queue then stores the waiting times corresponding to each elementary event, which is used to propagate the state of the KMC simulation.
Notably, while the GT-KMC occurrence frequencies are closely related to the rates of the individual processes, they are certainly different, as GT-KMC takes into account the cooperation and competition between the different processes as well as the effect of lateral interactions between H adatoms on the reaction rates and offers insights from a different perspective. 

\begin{figure}[!ht]
    \centering
    \includegraphics[width=16cm]{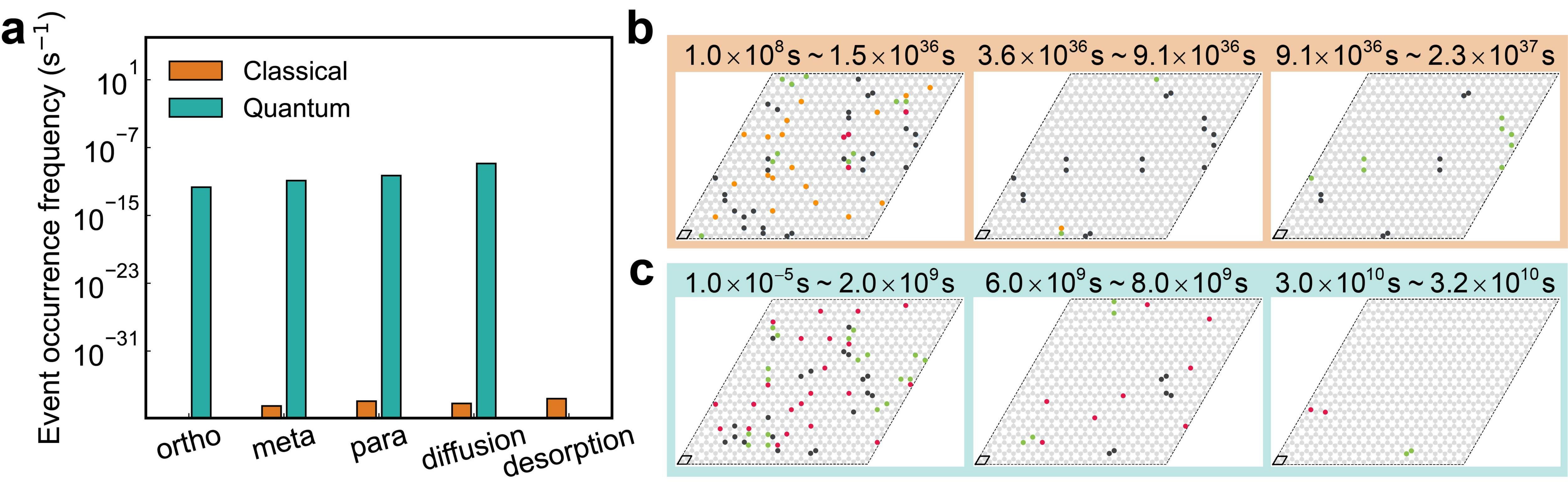}
    \caption{\textbf{
    Event occurrence frequencies of various processes from GT-KMC simulations.} 
    (a) Comparison of the classical and quantum event occurrence frequencies of H diffusion, desorption, and recombination on graphene at 100 K\@.
    Representative windows of KMC simulations at 100 K for (b) the classical scenario and (c) the quantum scenario. 
    Note that different timescales are displayed in the two cases.
    For each window, H atoms that diffuse are coloured red.
    H atoms that desorb or recombine as H$_2$ are coloured orange and green, respectively. 
    Immobile H atoms are coloured as black. 
    }
    \label{figure3}
\end{figure}

We focus on the low-temperature regime at 100 K\@.
For the ortho path, the instanton rate at 150 K is used in our KMC model for the quantum scenario and, as one can see from Figure~\ref{figure2}, the rate has become weakly temperature dependent by 150 K, a signature of deep tunnelling.
One can see from Figure~\ref{figure3} that the classical and quantum (instanton) scenarios differ drastically.
Classically, H atoms are effectively frozen on the graphene surface as diffusion and recombination processes have exceedingly low occurrence frequencies.
Quantum tunnelling significantly enhances the event occurrence frequency of these processes by $\sim$30 orders of magnitude.
The GT-KMC results also show that at such a low temperature, H atom desorption no longer occurs, even after accounting for quantum tunnelling. 
These findings are consistent with the classical and quantum rates discussed in the previous sections.
More importantly, the GT-KMC results confirm the qualitative difference in the role of the ortho pathway between the classical and quantum scenarios.
Classically, the ortho pathway is unfavourable due to the high potential energy barrier.
Multidimensional tunnelling effects dramatically boost the rate of the ortho process and the GT-KMC results show that this pathway becomes almost as important as the other two pathways in the quantum scenario.

To visualize the difference, we show snapshots from the GT-KMC simulations with the classical and quantum rates at 100 K in Figure~\ref{figure3}b,c (see the SI for details). 
In accordance with the discussion on the individual rates, KMC also predicts that, in the classical scenario, H atoms remain adsorbed over a long period of more than 1.5$\times$10$^{36}$ s (i.e.,\ longer than the age of the universe). For the few H atoms that do leave the surface, the desorption process dominates, whereas the H$_2$ formation processes are rare.
However, in the quantum scenario during a significantly shorter period (Figure~\ref{figure3}c) of 2.0$\times$10$^{9}$ s ($\sim$60 years, a relatively short period of time in interstellar space), we observe not only diffusion but also a large amount of H$_2$ formation processes (green dots). 
At the end of the simulation all H atoms have recombined. 

\begin{figure}[ht]
    \centering
    \includegraphics[scale=0.07]{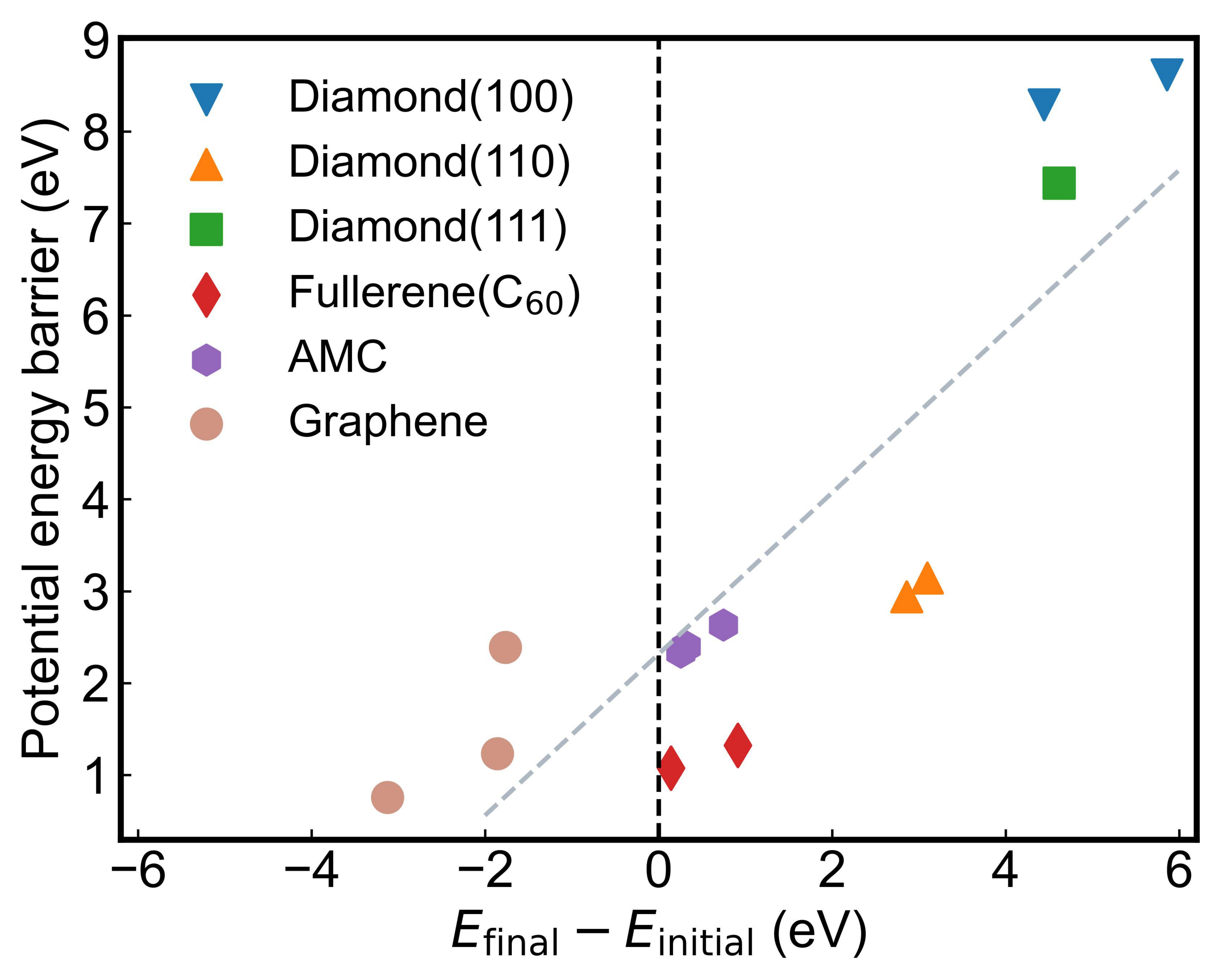}
    \caption{\textbf{H$_2$ formation on carbonaceous surfaces.}
    The vertical axis plots the 
    potential energy barriers for H$_2$ formation on several carbonaceous surfaces obtained by CI-NEB calculations.
    $E_{\rm final}-E_{\rm initial}$ is the energy difference between the final and the initial state of the formation process. The vertical dashed line separates the exothermic and the endothermic regimes, and the gray dashed line qualitatively describes the roughly linear correlation between the barrier and the energy difference.} 
    \label{figure4}
\end{figure}

In addition to graphene, other carbonaceous materials exist in nature in different forms; hence naturally they can also serve as substrates for H$_2$ formation. 
In general, carbon materials can exhibit sp$^2$ hybridization (e.g.,\ amorphous carbon and graphene) or sp$^3$ hybridization (e.g.,\ diamond).
For comparison, we calculate the potential energy barriers for H$_2$ formation on diamond, fullerene (C$_{60}$), and amorphous monolayer carbon (AMC) surfaces at different adsorption sites using the CI-NEB method (see Figure~\ref{figure4}). 
The detailed formation configurations are provided in SI Section VI. Overall, the energy barriers rise as the energy difference between the final and the initial state increases, qualitatively satisfying the Bell--Evans--Polanyi (BEP) principle \cite{chen_broadening_2021} which predicts a linear relation between the activation energy and the enthalpy of reaction. 
On the diamond surfaces, the associative H$_2$ desorption processes are endothermic by
more than 2 eV in each case, 
and correspondingly the barriers are at least 3 eV, meaning that H$_2$ formation is unfavourable on this type of surface.
We attribute this to the strong chemical binding of the hydrogen to sp$^3$ carbon surfaces.
On the AMC surfaces, all formation processes exhibit large barriers of more than 2.3 eV, 
which are close to the highest energy values on graphene, suggesting that H$_2$ formation via classical over-the-barrier processes on AMC is more difficult than on graphene.
It is however possible that H$_2$ formation on AMC could be enabled by quantum tunnelling, similarly to the case of the ortho pathway on graphene.
However, diffusion via tunnelling on AMC could be hampered by the fact that different sites on the surface have different binding energies, meaning that some thermal activation will be necessary. Therefore, H atom clusters might be difficult to form on AMC at low temperatures.
On fullerene (C$_{60}$) we found two pathways with relatively low barriers, one of which is even lower than the para pathway on graphene and is only slightly endothermic.
The potential energy curves of these processes are also shown in SI Section VI. The curvature at the barrier top is often a good indicator of the strength of quantum tunnelling. The pointier the barrier, the higher the crossover temperature and the stronger the tunnelling effects. We see that on other carbonaceous systems the barriers also have quite large curvatures. The large curvature can be attributed to three aspects: (i) the high energy cost for breaking C-H bonds, (ii) the short distance between the two adsorption sites of hydrogen and (iii) the strong C-C bonds that discourages large movement of C atoms during the transition.
Therefore, we can conclude that all the non-amorphous sp$^2$ carbon materials which we tested (i.e.,\ graphene and fullerene) are favorable substrates for H$_2$ formation.
%


To conclude, by extensively studying hydrogen formation processes on graphene, we revealed that the H$_2$ formation mechanisms of the quantum scenario are fundamentally different from those of the classical and quantum scenarios at low temperatures. 
It was widely believed that the LH mechanism for chemisorbed H atoms does not occur due to the low diffusion rates of H atoms adsorbed on the surface, which are thought to prevent long-range H diffusion.
We argue that such a picture is only true in the classical scenario.
With quantum mechanics, H diffusion can occur at the temperatures and timescales of the interstellar medium, as quantum tunnelling drastically accelerates the diffusion process but not the desorption process.
Classically, it is believed that the recombination of chemisorbed H atoms only occurs via the para site, and that recombination from the ortho site is impossible due to the high barrier.
Yet, quantum tunnelling enables the ortho recombination mechanism by enhancing the rate by tens of orders of magnitude, making this mechanism more competitive at low temperatures.
The tunnelling pathway for the ortho mechanism qualitatively differs from the classical pathway due to strong multidimensional quantum corner-cutting effects, featuring a concerted recombination mechanism instead of a stepwise one.

Theoretically, it is well known that DFT underestimates the barriers for covalent bond breaking or formation processes due to the self-interaction error.
Thus, as a first step toward the quantitatively exact treatment
we examine the influence of self-interaction error on our results by testing a hybrid functional, namely the HSE06 functional \cite{krukau_influence_2006} with D3 dispersion corrections \cite{grimme_consistent_2010}.  
%
%
%
As discussed in detail in SI Section VII, we found that
this change does not have any qualitative impact on our conclusions, and quantitatively it brought the rate of the ortho path even closer to the rate of the para path, further strengthening the finding that the ortho path is important in the quantum scenario.
%
%
We also note that our theoretical discussions are within the Born-Oppenheimer approximation and have not included non-adiabatic effects. According to previous studies, the dissociation of the C-H bond is accompanied by very large excitation energies\cite{bangRegulatingEnergyTransfer2013}. Therefore, we do not expect significant non-adiabatic effects for hydrogen recombination, diffusion and desorption processes at low temperatures without external fields.

This work highlights the pivotal impact quantum-mechanical effects can have for surface processes involving hydrogen, which casts new light on the understanding of H$_2$ formation on graphene and other carbonaceous surfaces.
%
%
Although graphene/graphite and their hydrogenated samples are among the most well controlled samples which can be nicely investigated with various types of experimental instruments, such measurements have been very rare due to the difficulty in detecting hydrogen and the misconception of the mechanism and speed of H$_2$ formation.
Here, an important fact revealed is that the quantum rates of H$_2$ formation are within the detectable regimes of many laboratory experimental techniques such as helium and neutron scattering, surface probing, and optical diffraction. 
Combined with the possibility of visualizing the position and vibration of hydrogen using advanced microscopies, experimental studies of H$_2$ formation on graphene/graphite are desirable to compare with our theoretical predictions.
Last but not the least, H$_2$ formation on other materials may also feature strong tunnelling effects, and we urge experimental works to take tunnelling effects into account in their data analysis. 
\begin{acknowledgement}

The authors thank X.-Z. Li and Y.-X. Feng for helpful discussions on this topic.
This work was supported by the Strategic Priority Research Program of Chinese Academy of Sciences under Grant No. XDB33000000, the National Key R\&D Program of China under Grant No. 2021YFA1400500, and the National Natural Science Foundation of China under Grant No. 11974024, 92165101.
We are grateful for computational resources provided by Peking University, the TianHe-1A supercomputer, Shanghai Supercomputer Center, and Songshan Lake Materials Lab.

\end{acknowledgement}

\begin{suppinfo}

The Supporting Information is available free of charge.
\begin{itemize}
  \item Details of calculations and detailed results of this work; Analysis of the quantum and classical pathways for hydrogen diffusion, recombination, and desorption; Minimum energy pathway based 1D tunnelling corrections; H/D isotope effects; KMC simulation setup and convergence tests; Geometries for hydrogen molecule formation on carbonaceous surfaces; Tests with a hybrid exchange correlation functional.

\end{itemize}

\end{suppinfo}

\bibliography{ref.bib}

\begin{mcitethebibliography}{62}
\bibitem[Not()]{Note-1}
Note that the true desorption rate will be even slower than that predicted by
  instanton theory. This is because instanton theory only models the transition
  to the physisorbed state and H atoms that reach the physisorbed state via
  tunnelling do not necessarily have the energy to escape the physisorption
  well.\relax
\mciteBstWouldAddEndPunctfalse
\mciteSetBstMidEndSepPunct{\mcitedefaultmidpunct}
{}{\mcitedefaultseppunct}\relax
\EndOfBibitem
\end{mcitethebibliography}

\end{document}



\title{Supplementary Information: Quantum Tunnelling Driven H$_2$ Formation on Graphene}

\author{Erxun Han}
\thanks{These authors contributed equally to this work}
\affiliation{School of Physics, Peking University, Beijing 100871, China}
\affiliation
{Interdisciplinary Institute of Light-Element Quantum Materials and Research Center for Light-Element Advanced Materials, Peking University, Beijing 100871, People's Republic of China}
\author{Wei Fang}
\thanks{These authors contributed equally to this work}
\affiliation{State Key Laboratory of Molecular Reaction Dynamics and Center for Theoretical Computational Chemistry, Dalian Institute of Chemical Physics, Chinese Academy of Sciences, Dalian 116023, P. R. China.}
\affiliation{Department of Chemistry, Fudan University, Shanghai 200438, China}
\affiliation{Laboratory of Physical Chemistry, ETH Zurich, CH-8093 Zurich, Switzerland}
\author{Michail Stamatakis}
\affiliation{Thomas Young Center and Department of Chemical Engineering, University College London, Torrington Place, London WC1E 7JE, United Kingdom}
\author{Jeremy O. Richardson}
\affiliation{Laboratory of Physical Chemistry, ETH Zurich, CH-8093 Zurich, Switzerland}
\author{Ji Chen}
\email{ji.chen@pku.edu.cn}
\affiliation{School of Physics, Peking University, Beijing 100871, China}
\affiliation
{Interdisciplinary Institute of Light-Element Quantum Materials and Research Center for Light-Element Advanced Materials, Peking University, Beijing 100871, People's Republic of China}
\affiliation{Frontiers
Science Center for Nano-Optoelectronics, Peking University, Beijing 100871, People's Republic of China}
\affiliation
{Collaborative Innovation Center of Quantum Matter, Beijing 100871, People’s Republic of China}

\renewcommand{\figurename}{\textbf{Supplementary Figure}}
\renewcommand{\tablename}{\textbf{Supplementary Table}}



\maketitle
\thispagestyle{empty}
\tableofcontents

\clearpage
\setcounter{page}{1}

\section{Details of calculations and detailed results of this work}
\begin{figure}[!ht]
    \centering
    \includegraphics[scale=0.8]{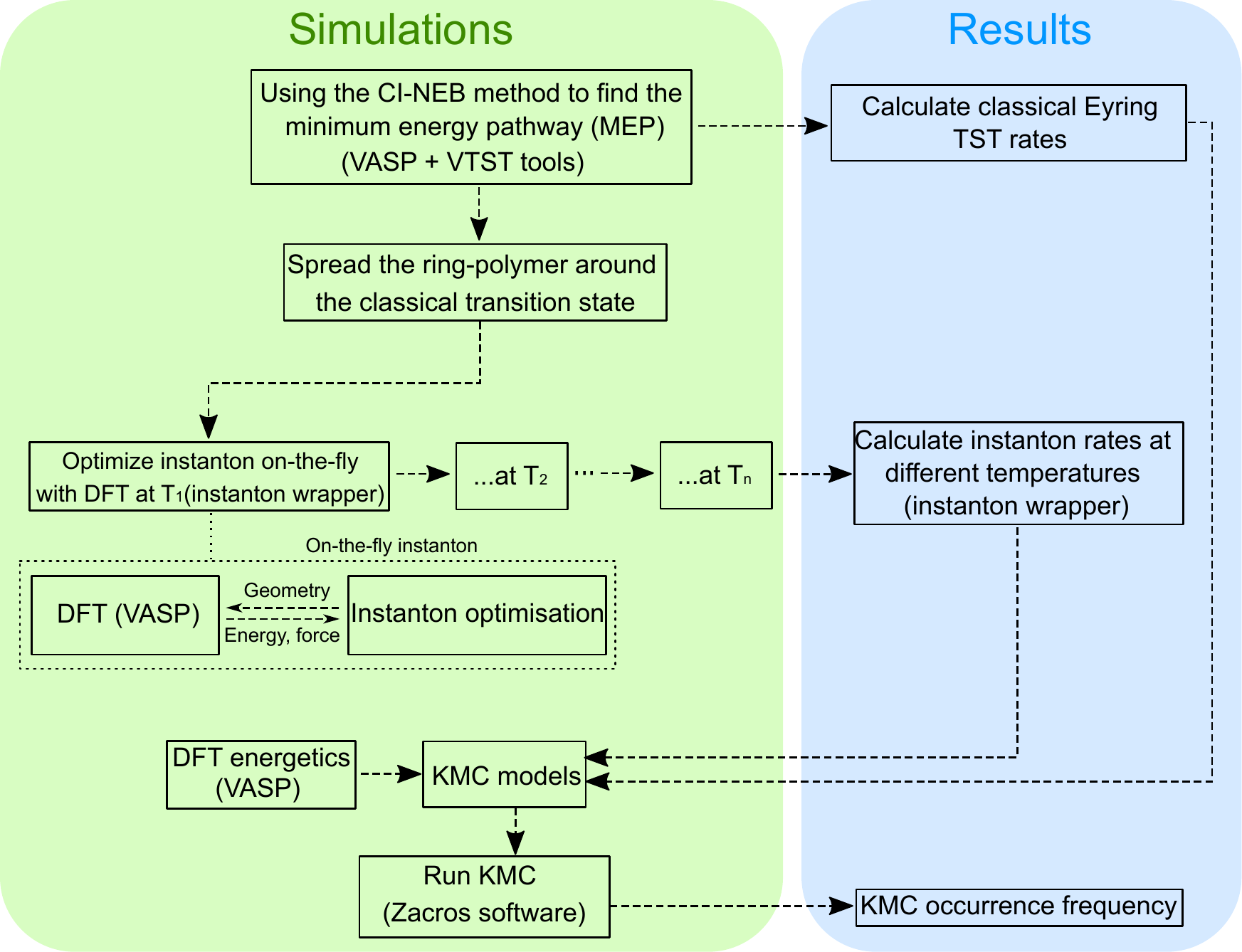}
    \caption{Workflow diagram of calculations. 
    We start with the classical transition state theory (TST) rate calculations for the different processes that can occur in the graphene-hydrogen system.
    Climbing image nudged elastic band (CI-NEB) calculations were performed using VASP \cite{kresse_efficient_1996} with VTST tools \cite{henkelman_climbing_2000}.
    For the quantum mechanical rates, \textit{ab initio} instanton calculations were performed at different temperatures for each process.
    The instanton optimisations for each process are performed in the order from the highest temperature ($T_1$) to the lowest temperature ($T_n$), and the optimised instanton at $T_k$ would be used as the initial guess for the instanton optimisation at $T_{k+1}$.
    Finally, occurrence frequencies of each process were obtained from kinetic Monte Carlo (KMC) calculations preformed with Zacros \cite{stamatakis_graph-theoretical_2011}.
    The KMC models were built from DFT results, classical rate results, and instanton rate results.
    }
    \label{fig5}
\end{figure}

First we provide a overall workflow describing how the different simulation methods are used and combined together in this work (Fig.~\ref{fig5}).

\noindent\textbf{Density functional theory.}
Our density-functional theory calculations were carried out using the Vienna \textit{ab initio} simulation package (VASP) \cite{kresse_efficiency_1996,kresse_efficient_1996}.
The Perdew-Burke-Ernzerhof (PBE) \cite{perdew_generalized_1997} exchange-correlation functional was used along with the D3 correction \cite{grimme_consistent_2010} to account for van der Waals interactions. 
A plane-wave cut off of 400 eV was used, and spin polarisation was enabled for the adsorption/desorption calculations. 
The graphene was modelled using an one-layer slab in a 4$\times$4 supercell with a 2$\times$2$\times$1 $k$-point mesh. 
A vacuum of at least 12.8~\AA\ was placed above each slab (20.8 \AA~for desorption process). 
The climbing image nudged elastic band (CI-NEB) method \cite{henkelman_climbing_2000} was used to obtain the potential energy barriers and minimum energy pathways (MEP) in mass-weighted coordinates. 
The force convergence criteria for geometry optimisations and CI-NEB calculations was 0.02 eV$\cdot$\AA$^{-1}$.
%
%
As shown in Supplementary Table~\ref{tab8}, the potential energy barriers of H$_2$ formation processes change the barriers by less than 2\% when the plane-wave cutoff or the $k$-point mesh size were increased.
This indicates that our conclusions are insensitive to the corresponding settings.
\begin{table}
\caption{
Convergence tests on potential energy barriers of H$_2$ formation processes with respect to DFT settings: plane-wave cutoff energy and $k$-point mesh size. 
The structures are fixed at the optimised geometries.\\}
\label{tab8}
\begin{tabular}{l|c|c|c}
\hline\hline
\multicolumn{1}{c|}{\multirow{3}{*}{Process}} & \multicolumn{3}{c}{Potential energy barrier (eV)} \\ \cline{2-4} 
\multicolumn{1}{c|}{} & Cutoff energy (400 eV)  & Cutoff energy (600 eV) & Cutoff energy (400 eV)\\
\multicolumn{1}{c|}{} & \multicolumn{1}{c|}{$k$-point mesh ($2\times2\times1$)} & \multicolumn{1}{c|}{$k$-point mesh ($2\times2\times1$)} & \multicolumn{1}{c}{$k$-point mesh ($4\times4\times1$)}\\ \hline
Ortho & 2.389 & 2.397 & 2.372\\
Meta & 0.756 & 0.762 & 0.759 \\
Para & 1.234 & 1.237 & 1.258\\ \hline\hline
\end{tabular}
\end{table}

\noindent\textbf{Instanton rate theory.}
Quantum tunnelling rates and pathways were computed with ring-polymer instanton rate theory \cite{richardson_ring-polymer_2009,richardson_ring-polymer_2018} and extrapolated to the infinite-bead limit \cite{beyer_quantum_2016}. 
The instanton rate can be formally written as
\begin{equation}
    \label{eq1}
    k_{\rm inst}(\beta;N) = A_{\rm inst}(\beta;N) \, e^{-W(E_\text{I})/\hbar-\beta E_\text{I}},
\end{equation}
where $\beta=1/k_{\text{B}}T$ and $N$ is the number of beads used to discretise the instanton trajectory. The abbreviated action is defined as
\begin{equation}
    \label{eq2}
    W(E)=2\int_{\textbf{x}}\sqrt{2\left[V(q)-E\right]}~\text{d}q,
\end{equation}
where \textbf{x} is a given reaction path, $V(q)$ is the potential energy along the path and $q$ is the mass-weighted coordinate.
For the instanton rate, $W$ is computed for an instanton trajectory with energy $E_\text{I}$.
As is typical, in all our calculations there was a one-to-one correspondence between $\beta\hbar$ and the instanton energy $E_\text{I}$.
$A_{\rm inst}$ is a measure of the fluctuations around the instanton trajectory.
The multidimensional instantons were obtained via first-order saddle-point optimisations 
at different temperatures with the total force converged to below 0.02 eV$\cdot$\AA$^{-1}$. 
The potential energy surface was calculated on-the-fly with DFT, performed using a python wrapper. 
Detailed results of the instanton calculations are listed in Supplementary Table \ref{tab7}, and the number of beads was between 32 and 90 to converge the rate to a desired accuracy. 
%
%
%
%
%
Notably, we calculated the $N$-bead instanton rates at 100 K for H$_2$ formation on meta and para sites with 64 and 90 beads, respectively. 
In both cases, the difference between the two results is less than a factor of 2.
Considering the classical rates are many orders of magnitude smaller than the instanton rates, we are confident that the $N$-bead instanton rates used are sufficiently accurate. 
We also note that the $W$ action, which contributes to the exponential part of the instanton rate, barely changes going from 64 to 90 beads, which also indicates that the number of beads is converged.

\begin{table}
\caption{Details of multidimensional instanton calculations.\\}
\begin{tabular}{c|c|c|c|c|c}
\hline\hline
Process & Temperature (K) & $N$ bead instanton rate (s$^{-1}$) & Beads & $W/\hbar$ & $E_{\rm I}$ (eV) \\ \hline
\multirow{7}{*}{Ortho}
 & \multicolumn{1}{c|}{150} & 2.83$\times10^{-12}$ & 90 & 59.8 & 0.11 \\
 & \multicolumn{1}{c|}{150} & 2.69$\times10^{-12}$ & 64 & 59.8 & 0.10 \\
 & \multicolumn{1}{c|}{200} & 2.45$\times10^{-11}$ & 64 & 56.7 & 0.15 \\
 & \multicolumn{1}{c|}{300} & 1.35$\times10^{-9}$ & 64 & 47.0 & 0.37 \\
 & \multicolumn{1}{c|}{400} & 1.23$\times10^{-7}$ & 64 & 34.9 & 0.74 \\
 & \multicolumn{1}{c|}{500} & 1.97$\times10^{-5}$ & 32 & 22.6 & 1.21 \\
 & \multicolumn{1}{c|}{600} & 6.04$\times10^{-3}$ & 32 & 11.7 & 1.74 \\ \hline
\multirow{7}{*}{Meta} 
 & \multicolumn{1}{c|}{100} & 8.05$\times10^{-2}$ & 90 & 28.2 & 0.11\\
 & \multicolumn{1}{c|}{100} & 4.65$\times10^{-2}$ & 64 & 28.6 & 0.11 \\
 & \multicolumn{1}{c|}{150} & 1.64$\times10^{1}$ & 64 & 24.4 & 0.12 \\
 & \multicolumn{1}{c|}{200} & 4.69$\times10^{2}$ & 64 & 21.3 & 0.16 \\
 & \multicolumn{1}{c|}{250} & 6.86$\times10^{3}$ & 64 & 17.5 & 0.23 \\
 & \multicolumn{1}{c|}{300} & 8.56$\times10^{4}$ & 64 & 13.2 & 0.32 \\
 & \multicolumn{1}{c|}{350} & 1.16$\times10^{6}$ & 32 & 7.9 & 0.47 \\ \hline
\multirow{7}{*}{Para} 
 & \multicolumn{1}{c|}{100} & 5.15$\times10^{-10}$ & 90 & 41.3 & 0.17 \\
 & \multicolumn{1}{c|}{100} & 1.74$\times10^{-9}$ & 64 & 41.9 & 0.16 \\
 & \multicolumn{1}{c|}{150} & 5.57$\times10^{-7}$ & 64 & 38.9 & 0.20 \\
 & \multicolumn{1}{c|}{200} & 3.67$\times10^{-5}$ & 64 & 35.0 & 0.26 \\
 & \multicolumn{1}{c|}{250} & 9.92$\times10^{-4}$ & 64 & 29.2 & 0.36 \\
 & \multicolumn{1}{c|}{300} & 2.60$\times10^{-2}$ & 64 & 22.0 & 0.54 \\
 & \multicolumn{1}{c|}{350} & 8.19$\times10^{-1}$ & 32 & 13.5 & 0.76 \\ \hline
\multirow{5}{*}{Diffusion} 
 & \multicolumn{1}{c|}{100} & 5.08$\times10^{-10}$ & 64 & 51.8 & 0.07 \\
 & \multicolumn{1}{c|}{150} & 1.48$\times10^{-8}$ & 64 & 42.5 & 0.16\\
 & \multicolumn{1}{c|}{200} & 1.78$\times10^{-6}$ & 64 & 28.9 & 0.36\\
 & \multicolumn{1}{c|}{250} & 6.91$\times10^{-4}$ & 32 & 13.5 & 0.66\\ \hline
\multirow{3}{*}{Desorption} 
 & \multicolumn{1}{c|}{100} & 3.47$\times10^{-20}$ & 32 & 10.9 & 0.78 \\
 & \multicolumn{1}{c|}{150} & 1.43$\times10^{-10}$ & 32 & 6.6 & 0.84\\
 & \multicolumn{1}{c|}{180} & 5.38$\times10^{-7}$ & 32 & 2.8 & 0.89\\ \hline\hline
\end{tabular}
\label{tab7}
\end{table}
\clearpage

\noindent\textbf{Kinetic Monte Carlo simulations.} 
For kinetic Monte Carlo simulations of hydrogen formation we used the GT-KMC framework as implemented in the software Zacros \cite{stamatakis_graph-theoretical_2011,nielsen_parallel_2013,ravipati_caching_2020}.
Compared with previous KMC schemes, the GT-KMC framework represents the lattice, adlayer configurations, elementary events and lateral interactions using graphs. 
It also implements the cluster-expansion based models for the simulation of a wide range of processes. 
The energetics models were obtained by DFT calculations. 
The reaction mechanism includes the elementary events of H$_2$ formation, H-atom diffusion and desorption with prefactors and activation energies derived from transition-state theory (TST) rates for the classical scenario and from instanton rates for the quantum scenario. 
The pressure was zero throughout and the gas-phase fractions of H-atom and H$_2$ were both set to be zero. 
The lattice was constructed using a 20$\times$20 unit cell (800 sites), where
initial configurations were provided with 60 randomly distributed hydrogen atoms.
Statistical results were averaged over 20 simulations after equilibration.  
Details and convergence tests of our GT-KMC simulations are presented in section V.

\section{Analysis of the quantum and classical pathways for hydrogen diffusion, recombination, and desorption}
\label{sec1}

\begin{table}[!ht]
\centering
\caption{
Characteristic quantities of quantum tunnelling for H diffusion, recombination, and desorption processes on graphene.
$T_{\rm c}$ is the crossover temperature \cite{gillan_quantum-classical_1987,mills_generalized_1997} to quantum tunnelling.
$\Delta E_\text{t}$ is an effective activation energy reduction due to tunnelling, defined using the ratio between the instanton rate ($k_{\rm inst}$) and the Eyring transition state theory (TST) rate ($k_{\rm TST}$):  $\frac{1}{\beta}\ln\frac{k_{\rm inst}}{k_{\rm TST}}$ ($\beta=1/k_\text{B}T$).
$\Delta E_\text{t}$ values presented here are computed for the deep tunnelling cases: at 150 K for H$_2$ formation on ortho configuration, and at 100 K for the other processes.
The brackets show the percentage of $\Delta E_\text{t}$ with respect to the classical potential energy barrier.
The carbon atom tunnelling distance refers to the average distance the C atoms (to which the H atoms are adsorbed) travelled (one way) along the instanton path, computed for the deep tunnelling case.
The brackets show the percentage of the carbon tunnelling distance with respect to that computed for the minimum energy pathway (MEP). \\
}
\begin{tabular}{l|c|cc}
\hline\hline
Process  &  $T_{\rm c}$~(K) & $\Delta E_\text{t}$~(eV) & Carbon tunnelling distance (\AA) \\ \hline
Ortho  & 607 & 1.550 (77\%) & 0.15 (20\%)   \\
Meta & 413  & 0.250 (49\%) & 0.09 (13\%)  \\
Para  & 425 & 0.574 (57\%) & 0.11 (15\%)  \\
Diffusion & 286  & 0.453 (51\%) & 0.25 (32\%)  \\
Desorption  & 199  & 0.085 (12\%) & 0.13 (25\%) \\
\hline\hline
\end{tabular}
\label{tab1}
\end{table}

Supplementary Table~\ref{tab1} shows the high crossover temperature for quantum tunnelling and the large effective barrier reduction due to deep tunnelling, showing that H diffusion and recombination processes are quantum mechanical in nature.
Carbon tunnelling is also observed, albeit to a limited extent even in the deep tunnelling regime.

\begin{figure}
    \centering
    \includegraphics[width=16cm]{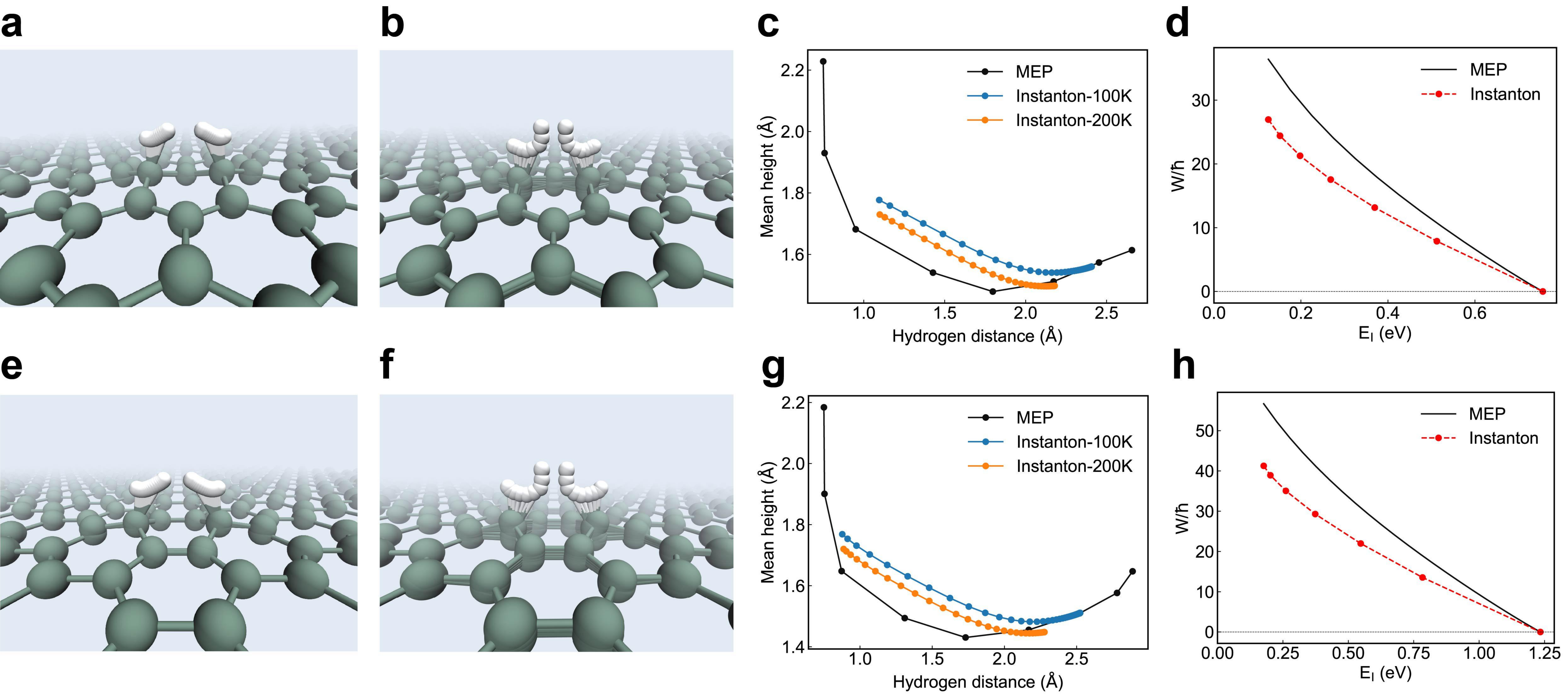}
    \caption{
    Comparison of the instanton trajectories and classical MEPs for H$_2$ formation on the meta and para configurations.
    Geometry of the instanton trajectory at 100 K (a) and classical MEP (b) for H$_2$ formation on the meta configuration. 
    (c) 2D representation (with the mean height of the two H atoms and the distance between them) of the classical MEP and instanton pathways at different temperatures. 
    (d) Comparison of the abbreviated action (Eq.~\ref{eq2}) calculated along the MEP and the instanton trajectories at different energies.
    (e)-(h) are the same as (a)-(d) but for H$_2$ formation on the para configuration at 100 K\@.\\
    }
    \label{fig1}
\end{figure}

Supplementary Fig.~\ref{fig1} shows obvious deviations of the instanton pathways from the MEPs, which is a sign of strong corner-cutting effects for the H recombination pathways on graphene.

\begin{figure}
    \centering
    \includegraphics[scale=0.085]{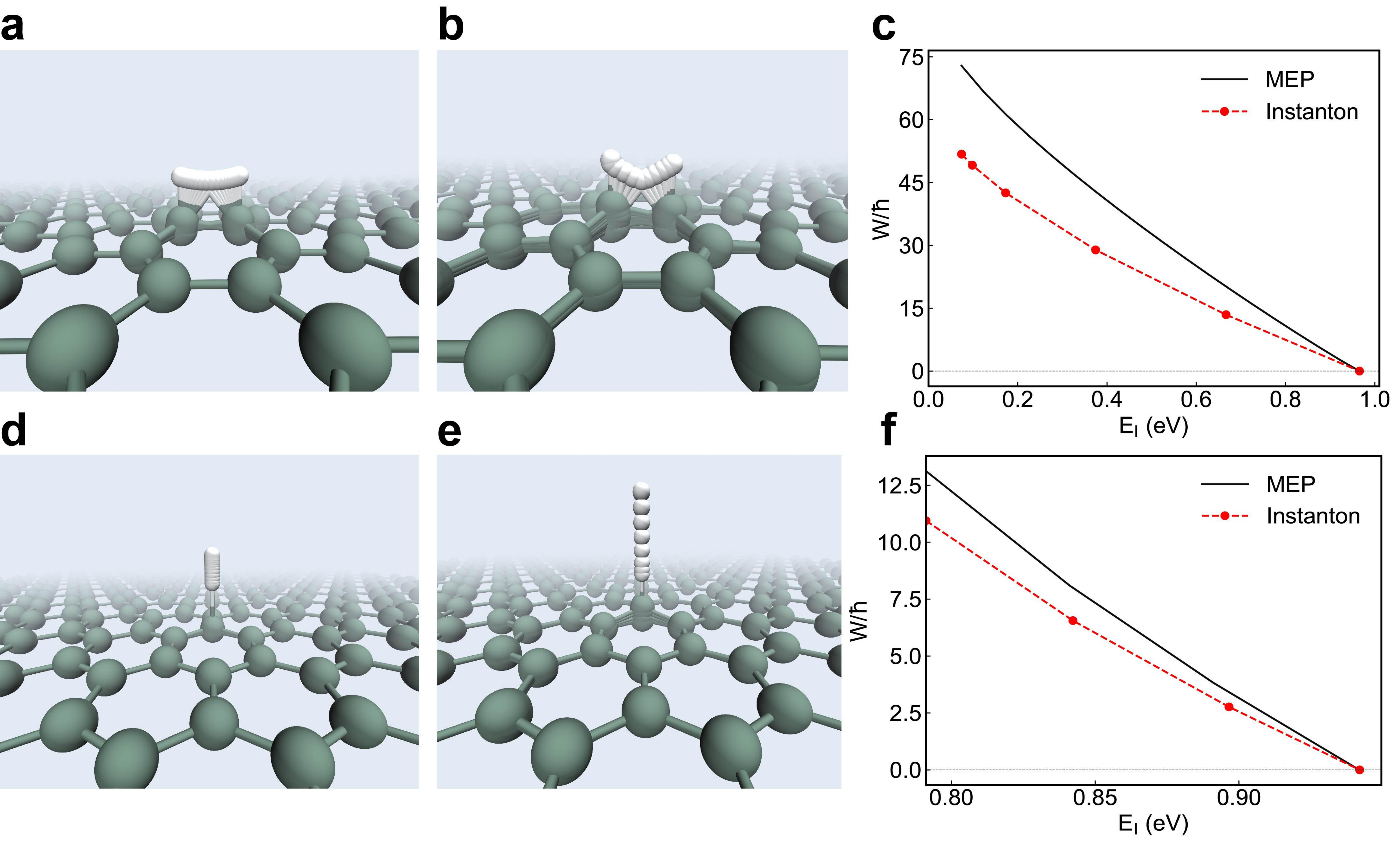}
    \caption{
    Same as Supplementary Fig.~\ref{fig1} for H diffusion (a-c) and desorption (d-f) processes.
    }
    \label{fig2}
\end{figure}

Supplementary Fig.~\ref{fig2} shows that strong corner-cutting effects also exist in H diffusion on graphene, but not so much for H desorption/adsorption processes.
For the desorption process, no tunnelling pathway exists below 0.79 eV, which is the energy difference between the initial and final states. 
This is because tunnelling cannot break energy conservation.

\begin{figure}
    \centering
    \includegraphics[scale=0.095]{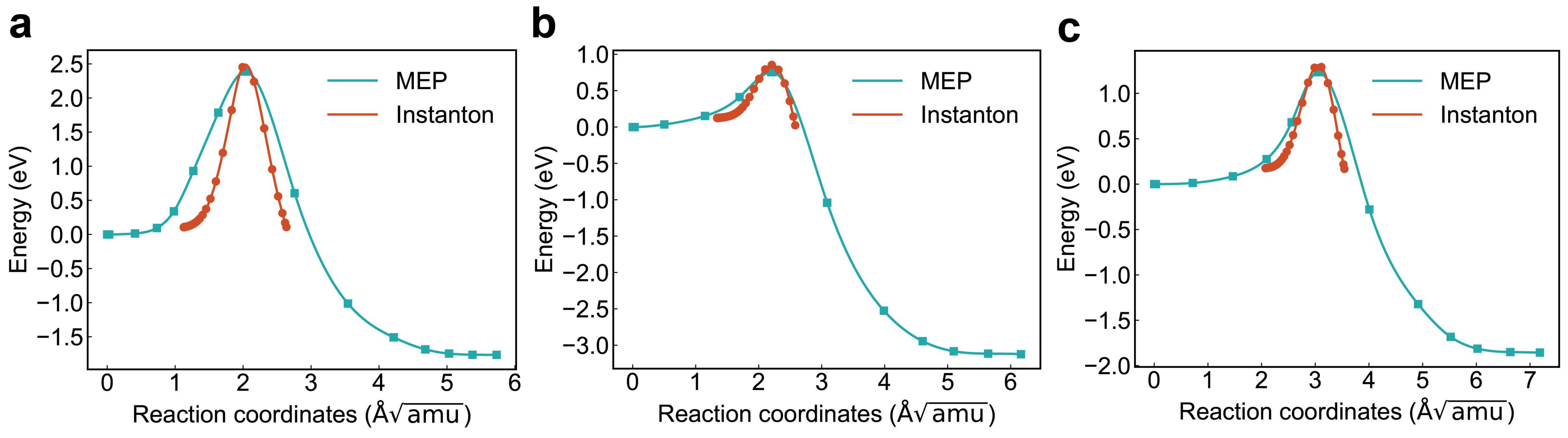}
    \caption{
    Comparison of MEP and instanton, 
    plotted along their respective mass-weighted reaction coordinates,
    for H$_2$ formation on (a) the ortho configuration at 150 K, (b) the meta configuration at 100 K and (c) the para configuration at 100 K. 
    The reaction coordinate for the instanton is shifted to align with the highest energy point of the MEP.
    }
    \label{fig3}
\end{figure}

Supplementary Fig.~\ref{fig3} shows that the deep tunnelling instantons take narrower pathways compared to the MEPs, especially for the ortho mechanism, which has the strongest corner-cutting effect among the three.
We also carefully verified that the ortho pathway does not have a ``symmetric" classical transition state, where the two H atoms move synchronously as in the instanton pathway.
Our climbing image nudged elastic band calculations did not find such pathway despite starting from a ``symmetric" initial guess.
Moreover, by taking the highest energy bead of the instanton for the ortho path at 600 K, in which the two H atoms are at the same height above the surface, and performing a frequency analysis, we found that this configuration has two imaginary frequencies.
In fact, the second imaginary mode (whose frequency is $\sim 750i$ cm$^{-1}$) is the asymmetric stretching of the two \ce{C-H} bonds.
Relaxation along this mode would ``break the symmetry" and result in the two H atoms at different heights.

\clearpage
\newpage

\section{Minimum energy pathway based 1D tunnelling corrections}
\label{sec2}
In this section we compare the one dimensional (1D) tunnelling factors computed on the MEPs using the WKB approximation \cite{Bell-1980} and the tunnelling factors from multi-dimensional instanton calculations.

The WKB transmission factor can be defined with an integral form or with a summation form.
%
%
The summation form is used in this analysis and we have tested that the integral form gives very similar results.
The WKB tunnelling factor ($\kappa_{\rm WKB}$), defined as the ratio between the WKB rate and the 1D transition state theory (TST) rate, is therefore given by:
\begin{equation}
\kappa_{\rm WKB}=\beta\hbar\tilde{\omega}_{\rm r}\mathrm{e}^{+\beta V^\ddagger} \sum_{n=0}^\infty \mathrm{e}^{-W(E_n)/\hbar}\mathrm{e}^{-\beta E_n},
\end{equation}
where $V^\ddagger$ is the energy of the transition state (TS) relative to reactants,
$\tilde{\omega}_{\rm r}$ is an effective reactant frequency,
$W(E)$ is the abbreviated action (Eq.~\ref{eq2}), and $\beta=1/k_\text{B}T$.
We note that there is no unique way of reducing a multi-dimensional system to a 1D model system, and thus there are many ways to define $\tilde{\omega}_{\rm r}$.
Here we define it using $\hbar\tilde{\omega}_{\rm r}/2=\Delta E_{\rm ZPE}$, where
$\Delta E_{\rm ZPE}$ is the zero point energy (ZPE) difference between the reactant and transition states, which is positive for all the processes studied in this work.
This definition is well suited to the separable approximation implied in the 1D based tunnelling corrections.
Also under this definition, the ZPE corrected barrier for the 1D model is the same as that of the full-dimensional system.
We have considered other definitions, all of which give lower $E_0$, and as a result, combining $\kappa_{\rm WKB}$ with the full dimensional Eyring TST rate would result in an unphysical behaviour where the rate increases with decreasing temperature (due to inclusion of tunnelling paths below the reactant ZPE).

%
%
%

\begin{table}[!ht]
\caption{
Comparison of tunnelling factors computed using 1D WKB approximation ($\kappa_\text{WKB}$) and multidimensional instanton theory at different temperatures.
The multidimensional instanton tunnelling factor is defined as $\frac{k_{\rm inst}}{k_{\rm TST}}$, where $k_{\rm inst}$ is the multidimensional instanton rate and $k_{\rm TST}$ is the multidimensional Eyring TST rate.\\
}
\begin{tabular}{l|cc|cc}
\hline\hline
\multicolumn{1}{l|}{\multirow{2}{*}{Process}} & \multicolumn{2}{c|}{1D WKB} & \multicolumn{2}{c}{Multidimensional instanton} \\ \cline{2-5} 
\multicolumn{1}{c|}{} & \multicolumn{1}{c}{100 K} & \multicolumn{1}{c|}{300 K} & \multicolumn{1}{c}{100 K} & \multicolumn{1}{c}{300 K} \\ \hline
Ortho & $6.70\times10^{33}$(150 K) & $1.02\times10^{3}$ & $1.33\times10^{44}$(150 K) & $1.21\times10^{12}$ \\
Meta & $1.64\times10^{16}$ & $1.89\times10^{1}$ & $3.70\times10^{12}$ & $5.94\times10^{0}$ \\
Para & $3.25\times10^{29}$ & $1.20\times10^{1}$ & $1.16\times10^{28}$ & $1.57\times10^{2}$ \\
Diffusion & $7.89\times10^{14}$ & $3.36\times10^{0}$ & $6.31\times10^{21}$ & - \\
Desorption & $7.93\times10^{2}$ & $2.07\times10^{0}$ & $5.88\times10^{4}$ & - \\ 
\hline\hline
\end{tabular}
\label{tab2}
\end{table}
Supplementary Table~\ref{tab2} shows a comparison of the tunnelling factor from 1D WKB approximation and the multidimensional instanton theory.
One can see that the 1D tunnelling factors are significantly underestimated for the recombination mechanisms from the ortho site as well as for H diffusion in the low temperature deep tunnelling regime.
Especially for the ortho pathway, the 1D WKB corrected rate is more than 10 orders of magnitude lower than the multidimensional instanton rate. 
Moreover, it is more than 12 orders of magnitude lower than the rate for the para pathway, suggesting that 1D tunnelling approximations predict that the ortho pathways is not important at all, which is qualitatively incorrect.
%
%

\begin{figure}
    \centering
    \includegraphics[width=8.5cm]{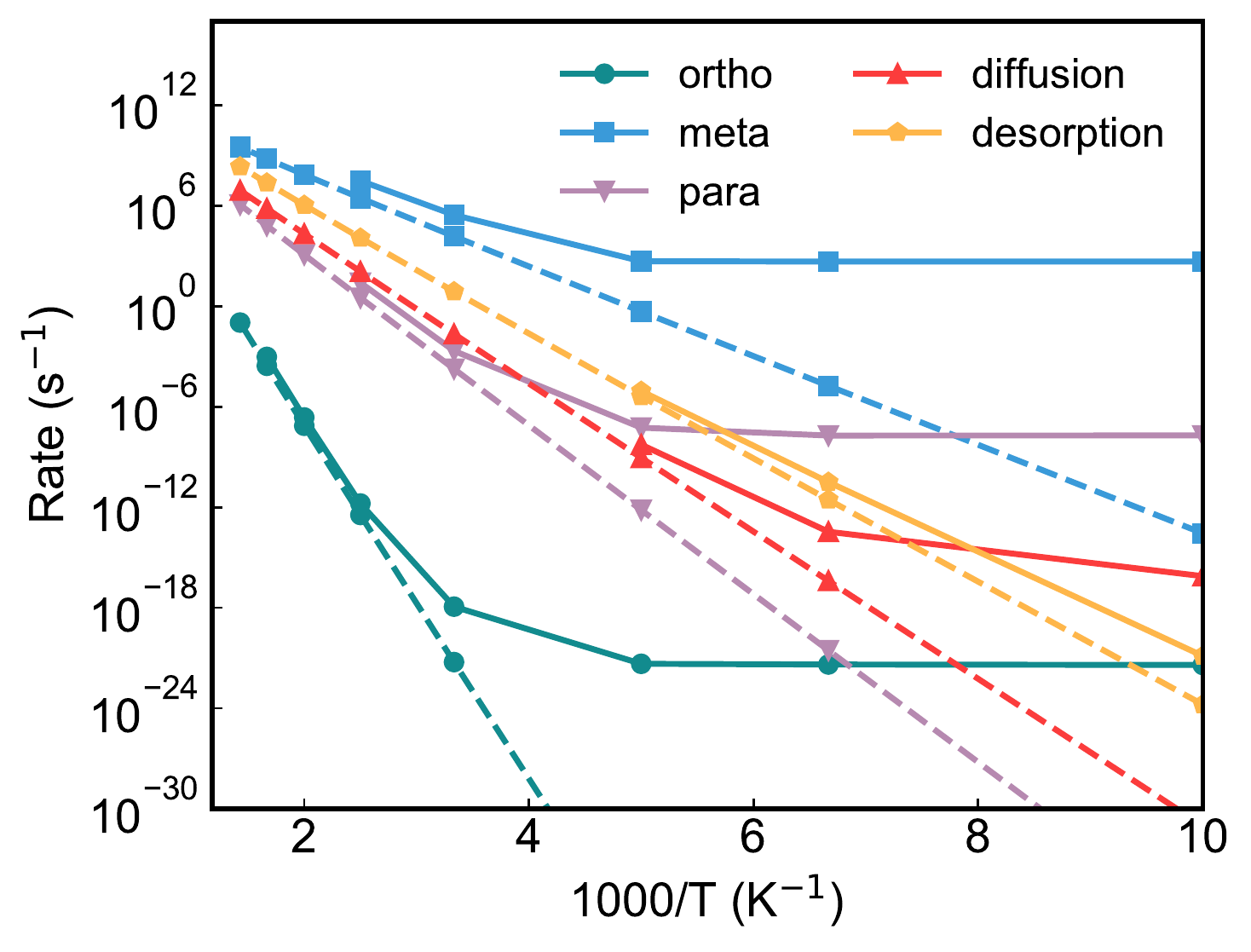}
    \caption{
    1D WKB corrected rates, defined as $\kappa_\text{WKB}k_\text{TST}$, for H atom processes on graphene (solid lines).
    The dashed lines represent the multidimensional Eyring transition state theory (TST) rates ($k_\text{TST}$).
    }
    \label{fig4}
\end{figure}

For the meta path, the tunnelling factor from 1D WKB correction is higher than that from multidimensional instanton theory.
One can see from Fig.~\ref{fig4} that the 1D WKB corrected rate for the meta path levels off at 200 K, much earlier than the instanton rates shown in the main text.
We expect that this is due to the neglect of non-separable effects, which results in an overestimation of the rate.
%

\clearpage
\section{H/D isotope effects}

The instanton trajectories to show the isotope effects are shown in Supplementary Fig.~\ref{fig14}.

\begin{figure}[!ht]
    \centering
    \includegraphics[width=11cm]{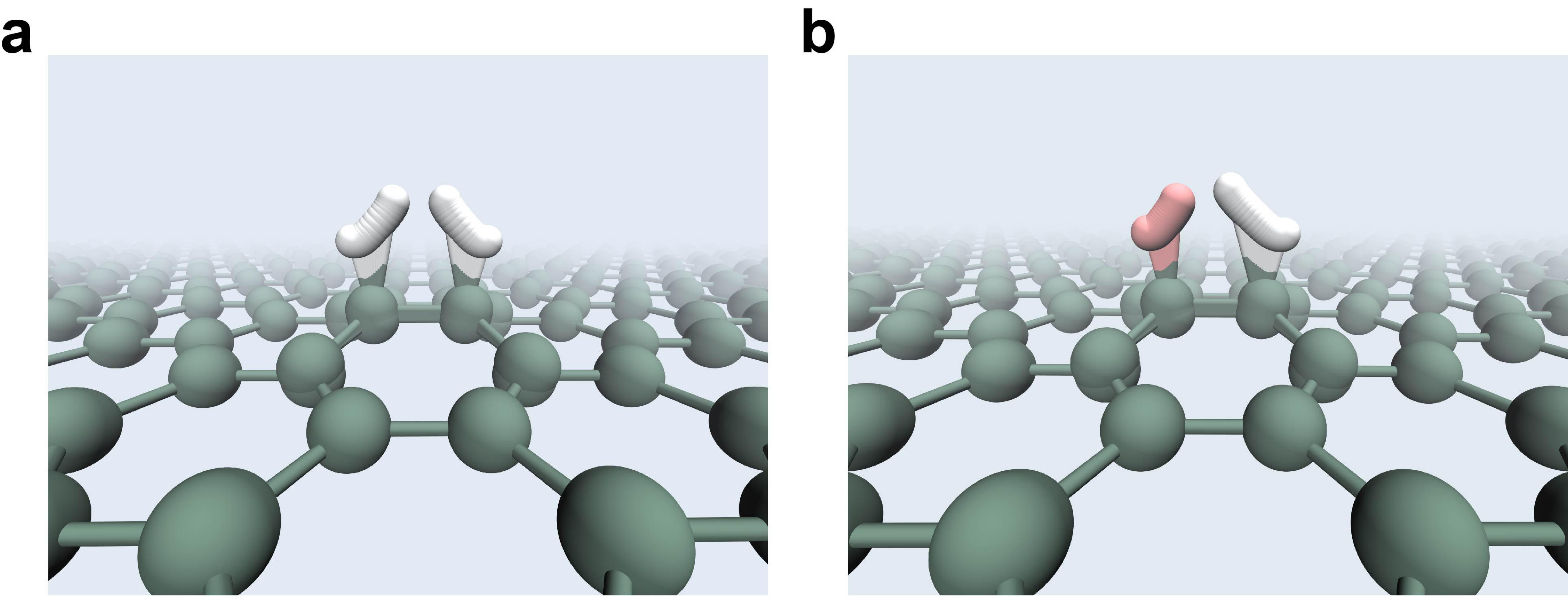}
    \caption{Instanton trajectories of (a) H$_2$ and (b) HD formation from ortho site at 150 K. The D atom is coloured red.}
    \label{fig14}
\end{figure}


\clearpage

\section{KMC simulation setup and convergence tests} \label{sec3}

For the energetics of the system in our kinetic Monte Carlo (KMC) simulations, we need to define the formation energy of different patterns of H atom clusters on the graphene substrate.
The cluster energy for a cluster of $n$ H atoms (whose geometry is denoted as $x^{(n)}$) is defined in a many-body expansion fashion \cite{roberts2009applied} as:
\begin{equation}
    E_{\rm clu}^{(n)}(x^{(n)})=E_{\rm tot}(x^{(n)})-n\times E_{\rm H}-E_{\rm gra}-\sum^{n-1}_{i=1} \sum_{j^{(i)}} E_{\rm clu}^{(i)}(x^{(i)}_{j^{(i)}}) 
\end{equation}
where $E_{\rm tot}$ is the total energy of the system, $E_{\rm H}$ is the energy of a single H atom, $E_{ \rm gra}$ is the energy of the graphene substrate (geometry optimised).
The index $j^{(i)}$ runs over all possible $x^{(i)}$ configurations (subclusters) contained in $x^{(n)}$.
In the cluster expansion we considered clusters up to the H atom trimer, and the energies of larger clusters (which rarely appear) are approximated by the sum of the energies of the subclusters from which they are formed.
The cluster energies obtained from DFT calculations are given in Supplementary Table~\ref{Ecluster}.
%
Note that the H$_2$ formation on the meta configuration has a positive cluster energy, indicating instability of the meta configuration.

\begin{table}[!ht]
\centering
\caption{
Cluster energies for different H adsorption clusters on graphene. 
Trimer-1 is the configuration where the 3 H atoms chemisorb on adjacent C atoms in one hexagonal ring on graphene.
Trimer-2 is the configuration where two H atoms adsorb on the para configuration on one hexagonal ring, and the third H atom adsorbs on one of the sites between them.\\
}
\begin{tabular}{c|c|ccc|cc}
\hline\hline
Cluster & Monomer & Dimer-ortho & Dimer-meta & Dimer-para & Trimer-1 & Trimer-2 \\ \hline
$E_{\rm clu}^{(n)}$ (eV) & $-$0.80 & $-$1.23 & 0.12 & $-$1.15 & 0.57 & 0.88 \\ 
\hline\hline
\end{tabular}
\label{Ecluster}
\end{table}


The Zacros software which we use for the KMC simulations takes as input activation energies and prefactors, which it automatically combines to give the rate constants.
The classical activation energy ($E^{\rm cl}_{\rm act}$) is defined as: 
\begin{equation}
    E_{\rm act}^{\rm cl}=V^{\ddagger}-\Delta E_{\rm ZPE}  
    \label{eq4}
\end{equation}
%
Note that the term ``classical'' here means neglecting quantum tunnelling but ZPE corrections are included.
The prefactor in the KMC simulations is accordingly given by 
\begin{equation}
A=k_{\rm TST} \, \mathrm{e}^{\beta E_{\rm act}^{\rm cl}}, 
\label{A}
\end{equation}
In the quantum scenario, there is no unique definition of the ``activation energy" like in the classical scenario.
However, as the code requires it, we convert the instanton rates into an effective activation energy with $k_\text{inst}(\beta)\equiv A \, \mathrm{e}^{-\beta E_{\text{eff-act}}^{\rm inst}(\beta)}$ (with the same $A$ as in Eq.~\ref{A}) for KMC simulations in the quantum scenario.
%
The simulation result depends only on the rate constant itself and is of course not affected by how we choose to separate the rate into the prefactor and activation energy.

The pressure was zero throughout and the gas-phase fractions of H-atom and H$_2$ were both set to be zero, which avoids numerical problems when dealing with extremely low partial pressures. Instead, we include pre-adsorbed H atoms on surface in KMC simulations and study the collective processes by varying the number of graphen sites and the number of randomly adsorbed H atoms. A typical setting consists of a 20$\times$20 supercell of graphene (800 sites) with 60 randomly distributed hydrogen atoms.

\begin{table}[!ht]
\caption{
Setup for KMC simulations at 100 K\@.
Classical (Eyring TST rates) and quantum (instanton rates extrapolated to the infinite bead limit) rates used in the KMC simulations.
For the 100 K quantum KMC simulations, we used the instanton rate at 150 K for the ortho path as an upper-bound approximation.
Such an approximation is valid because the instanton rate for this path becomes only weakly temperature dependent at 150 K and below.
The classical activation energy and prefactor are defined in Eqs. \ref{eq4} \& \ref{A} respectively.
\\
}
\begin{tabular}{l|c|c|c|c}
\hline\hline
Process & Classical rate (s$^{-1}$) & Instanton rate (s$^{-1}$) & \begin{tabular}[c]{@{}c@{}}Classical activation\\ energy (eV)\end{tabular} & Prefactor (s$^{-1}$) \\ \hline
Ortho & $9.00\times10^{-91}$ & $3.25\times10^{-12}$ & 2.04 & $7.92\times10^{11}$ \\
Meta & $2.78\times10^{-14}$ & $1.03\times10^{-1}$ & 0.51 & $9.53\times10^{11}$ \\
Para & $5.89\times10^{-38}$ & $6.82\times10^{-10}$ & 0.98 & $8.94\times10^{11}$ \\
Diffusion & $9.71\times10^{-32}$ & $6.13\times10^{-10}$ & 0.86 & $1.14\times10^{12}$ \\
Desorption & $1.67\times10^{-24}$ & $9.82\times10^{-20}$ & 0.72 & $1.97\times10^{12}$ \\
\hline\hline
\end{tabular}
\label{tab4}
\end{table}



\begin{figure}[!ht]
    \centering
    \includegraphics[scale=0.09]{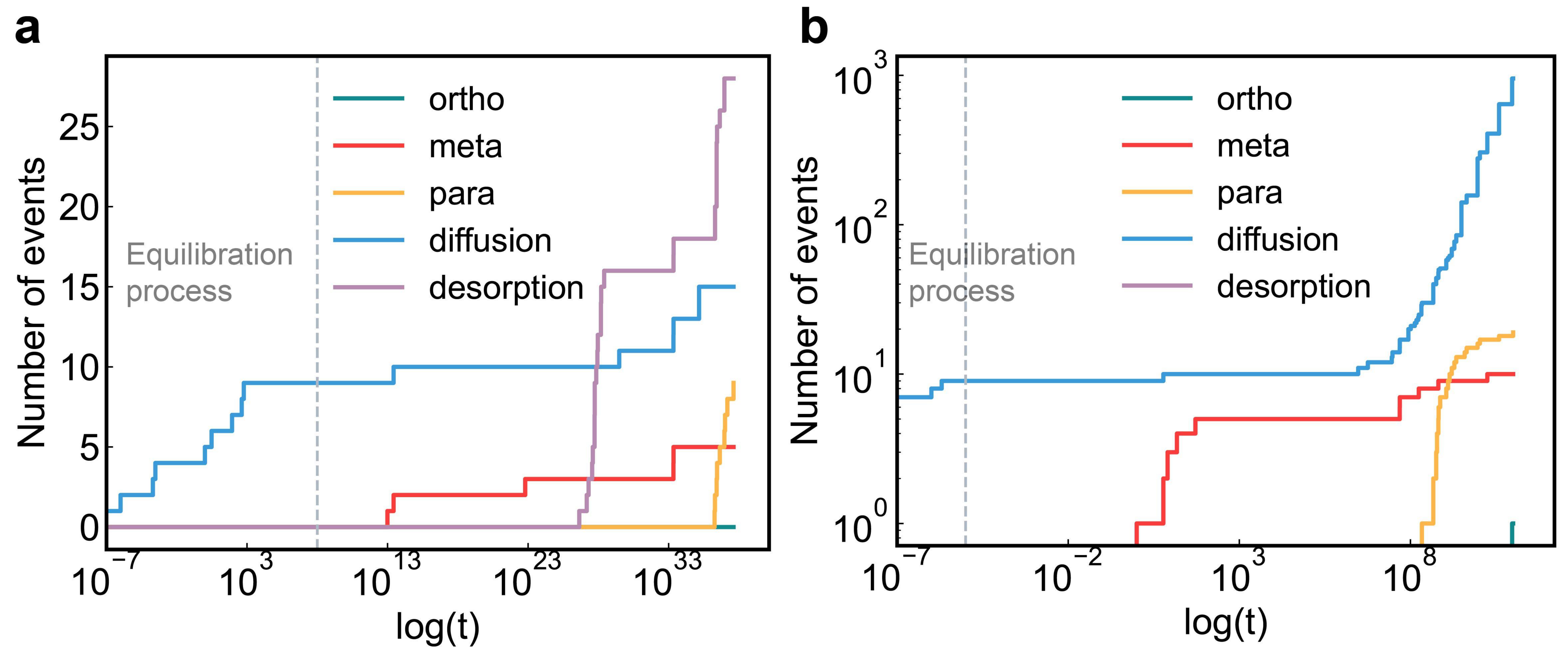}
    \caption{
    Cumulative event occurrence number for each process with respect to the simulation time for one GT-KMC trajectory in (a) the classical scenario and (b) the quantum scenario at 100 K.
    Note that the $y$ axis in (a) is in linear scale, as the number of events that occurred in the classical scenario is small.
    The grey dashed line marks the equilibration time, which we define as $1/10^{5}$ the timescale of the fastest process (i.e. the process with the highest rate in Table~\ref{tab4}).
    Events happened within the equilibrium time are discarded.
    }
    \label{fig13}
\end{figure}
Here we discuss more details of GT-KMC simulations results.
Fig.~\ref{fig13} shows a breakdown of when the events of each process occur during the GT-KMC simulation.
One can see that there is an initial equilibration process where a few ($\sim$10) diffusion events occur on a timescale several orders of magnitude smaller than the timescale of the fastest process.
Only events that occur after equilibration are counted in our results. 
One can see that the timescales at which each process occurs are consistent with the rates of the individual processes (Table~\ref{tab4}).
In the classical scenario, not many events occur over a long period of time due to fact that the individual processes happen at extremely slow rates in the classical scenario.
In fact the most frequently occurring process is H atom desorption, while H recombination is negligible.
In the quantum scenario however, thanks to deep quantum tunnelling, all the processes are significantly faster compared to the classical case.
The first process that occurs after equilibration is H recombination from the meta site, to be precise, from the meta dimer in a H atom trimer.
After this type of trimer is depleted, no H recombination from the meta site occurs for $\sim 10^8$ seconds.
Diffusion starts from $10^7$ seconds, recombination from the para site starts from $10^8$ seconds, and recombination from the ortho sites starts from $10^{10}$ seconds, in agreement with the rates of these processes.
No desorption events are observed in the quantum scenario simulations.
%

\begin{figure}[!ht]
    \centering
    \includegraphics[scale=0.55]{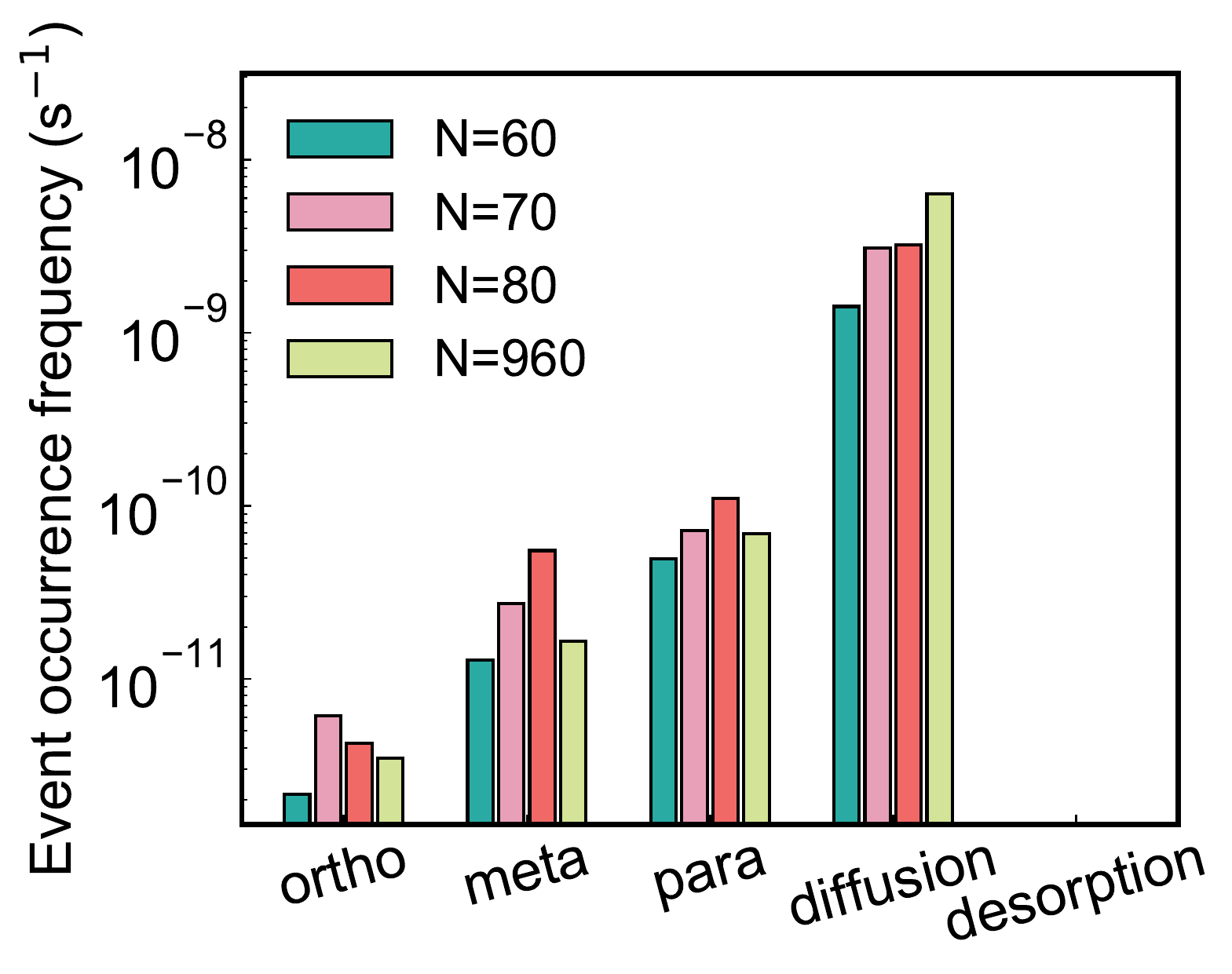}
    \caption{
    Size dependence of the event occurrence frequencies from KMC simulations. 
    N is the number of chemisorbed H atoms on the surface at the start of the simulation.
    The graphene slab size is 20$\times$20 (800 sites) for N = 60, 70, and 80.
    For N = 960 the system size is 80$\times$80 (12800 sites).
    }
    \label{fig6}
\end{figure}
We also tested the sensitivity of the KMC results with respect to the number of H atoms and the substrate size.
As shown in Supplementary Fig.~\ref{fig6}, we find that the KMC results are not sensitive to the system size.

\clearpage

\section{Geometries for hydrogen molecule formation on carbonaceous surfaces} \label{sec4}

%


\begin{figure}[!ht]
    \centering
    \includegraphics[width=\textwidth]{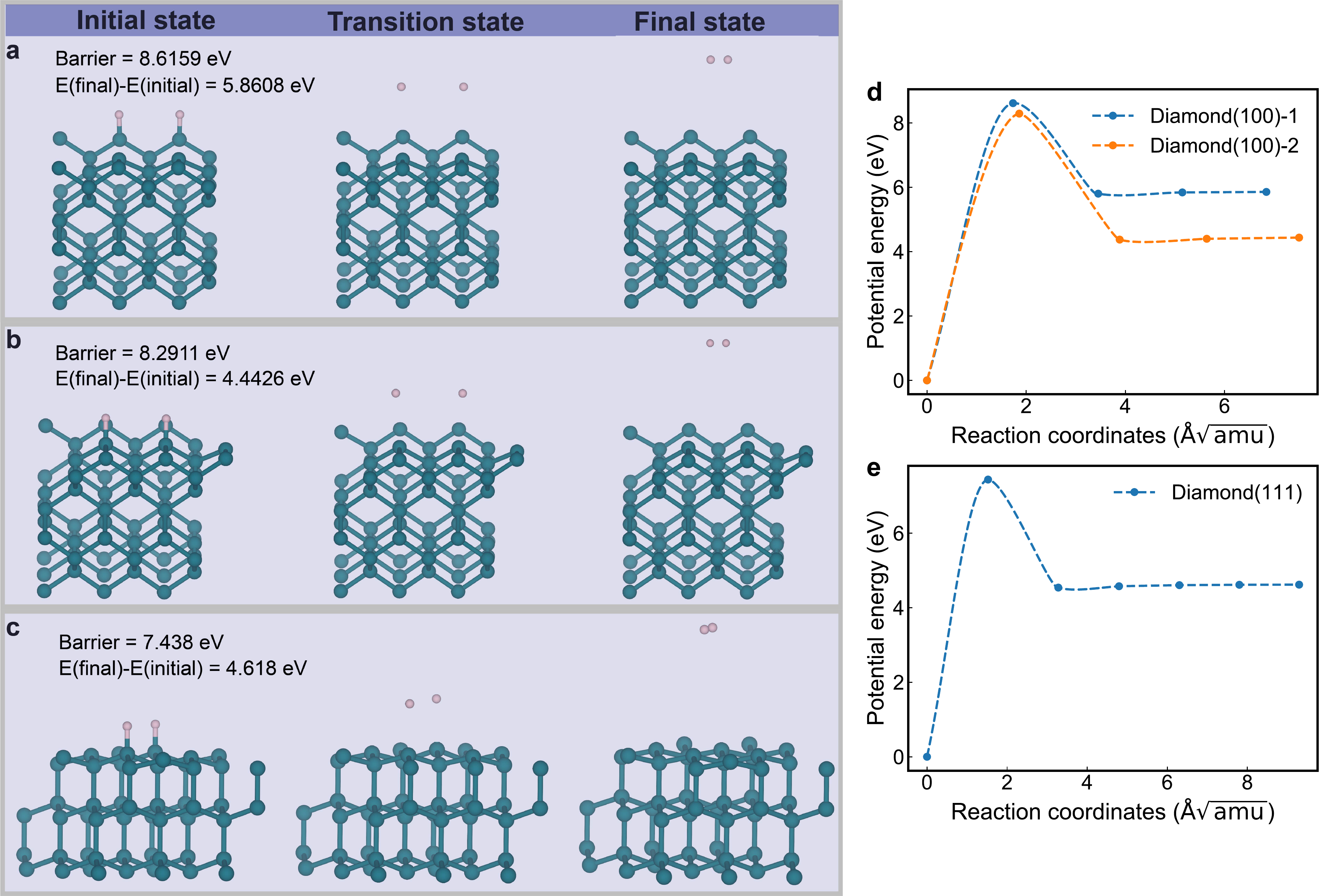}
    \caption{
    H$_2$ formation pathways on (a)-(b) diamond (100) surface and (c) diamond (111) surface at different adsorption sites.
    Potential energy profiles along the (mass-weighted) intrinsic reaction coordinate for H$_2$ formation on (d) diamond(100) and (e) diamond(111) surfaces. Filled symbols represent the NEB data and the dashed lines are a guide for the eye.
    $E$(final)$-E$(initial) is the relative energy of the final state with respect to the initial state in the given reaction.
    }
    \label{fig7}
\end{figure}

\begin{figure}[!ht]
    \centering
    \includegraphics[width=\textwidth]{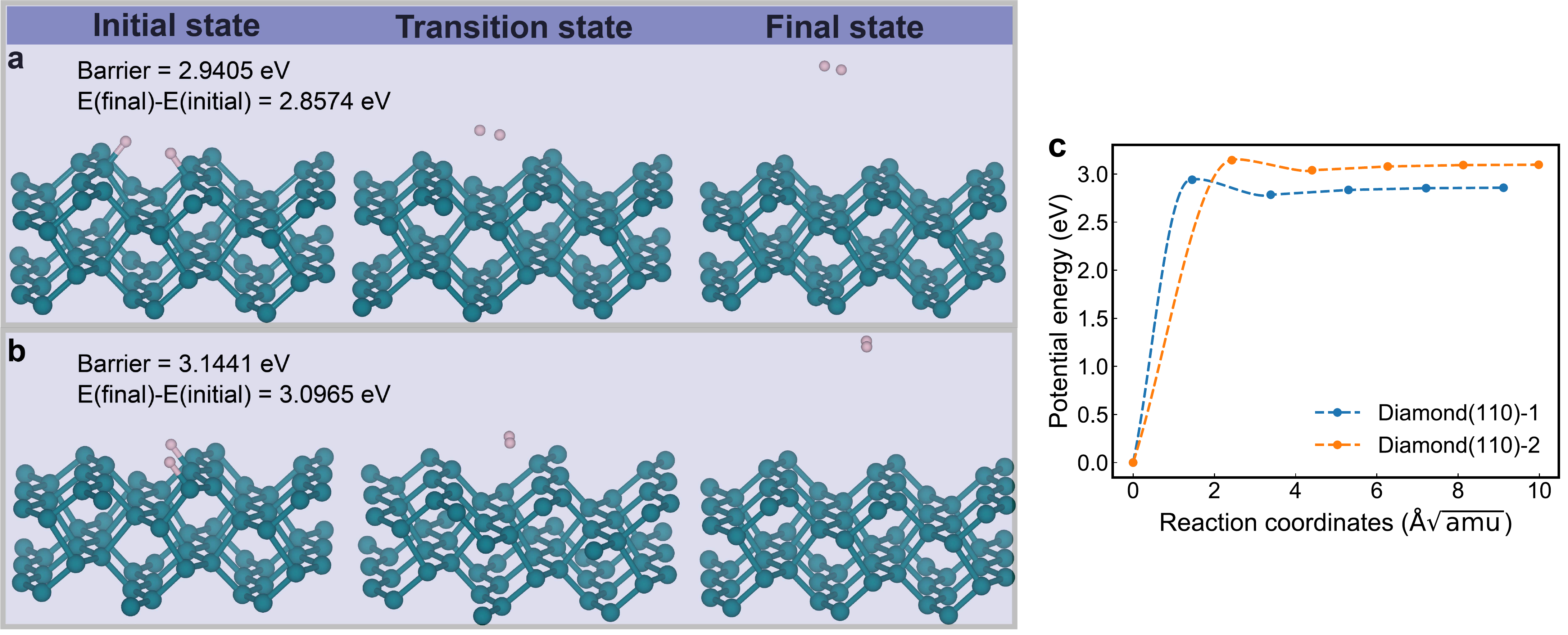}
    \caption{
    Same as Supplementary Fig.~\ref{fig7} for the diamond (110) surface.
    }
    \label{fig8}
\end{figure}

\begin{figure}[!ht]
    \centering
    \includegraphics[width=\textwidth]{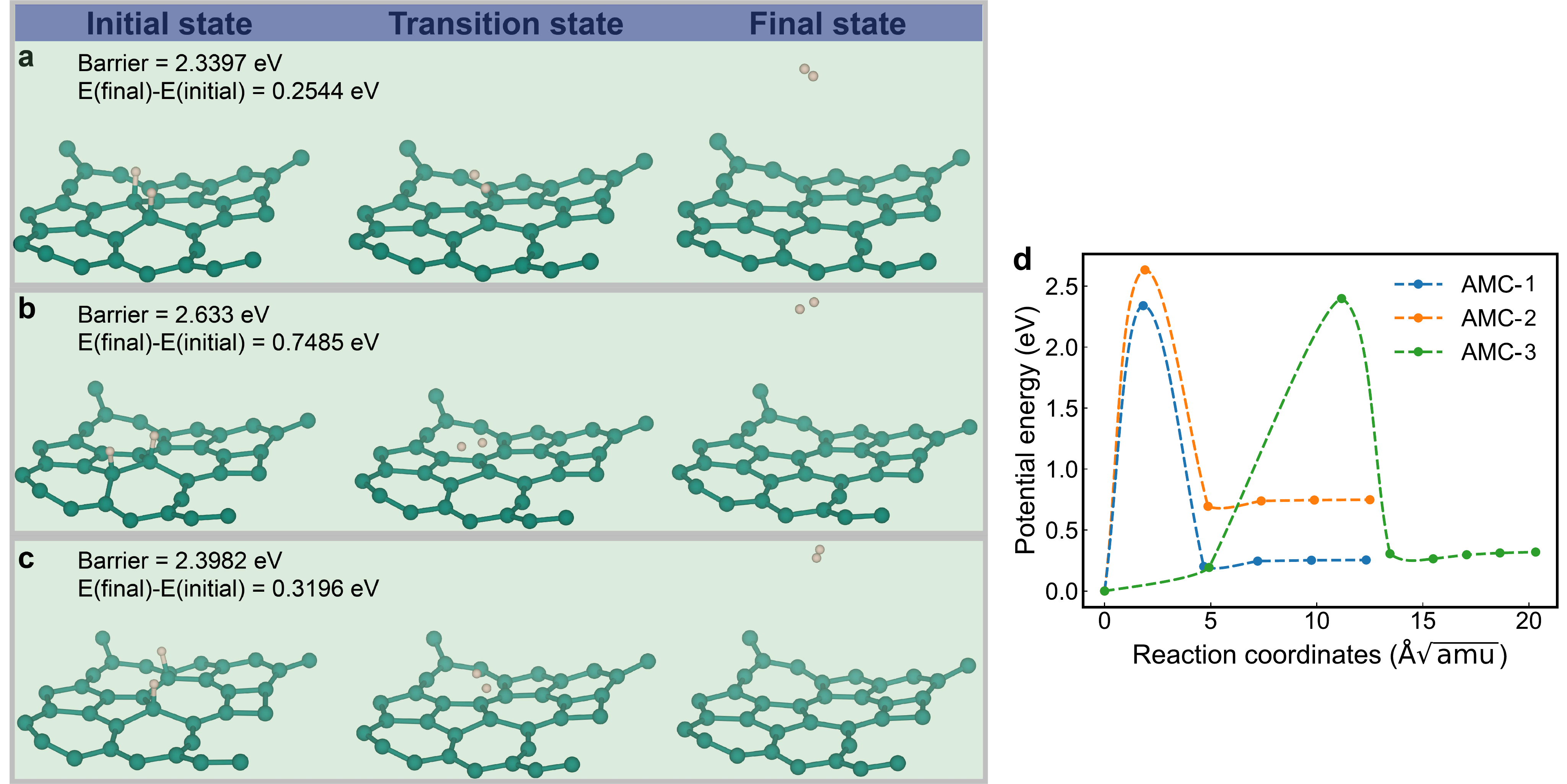}
    \caption{
    Same as Supplementary Fig.~\ref{fig7} for the amorphous monolayer carbon (AMC) surface.
    }
    \label{fig9}
\end{figure}

\begin{figure}[!ht]
    \centering
    \includegraphics[width=\textwidth]{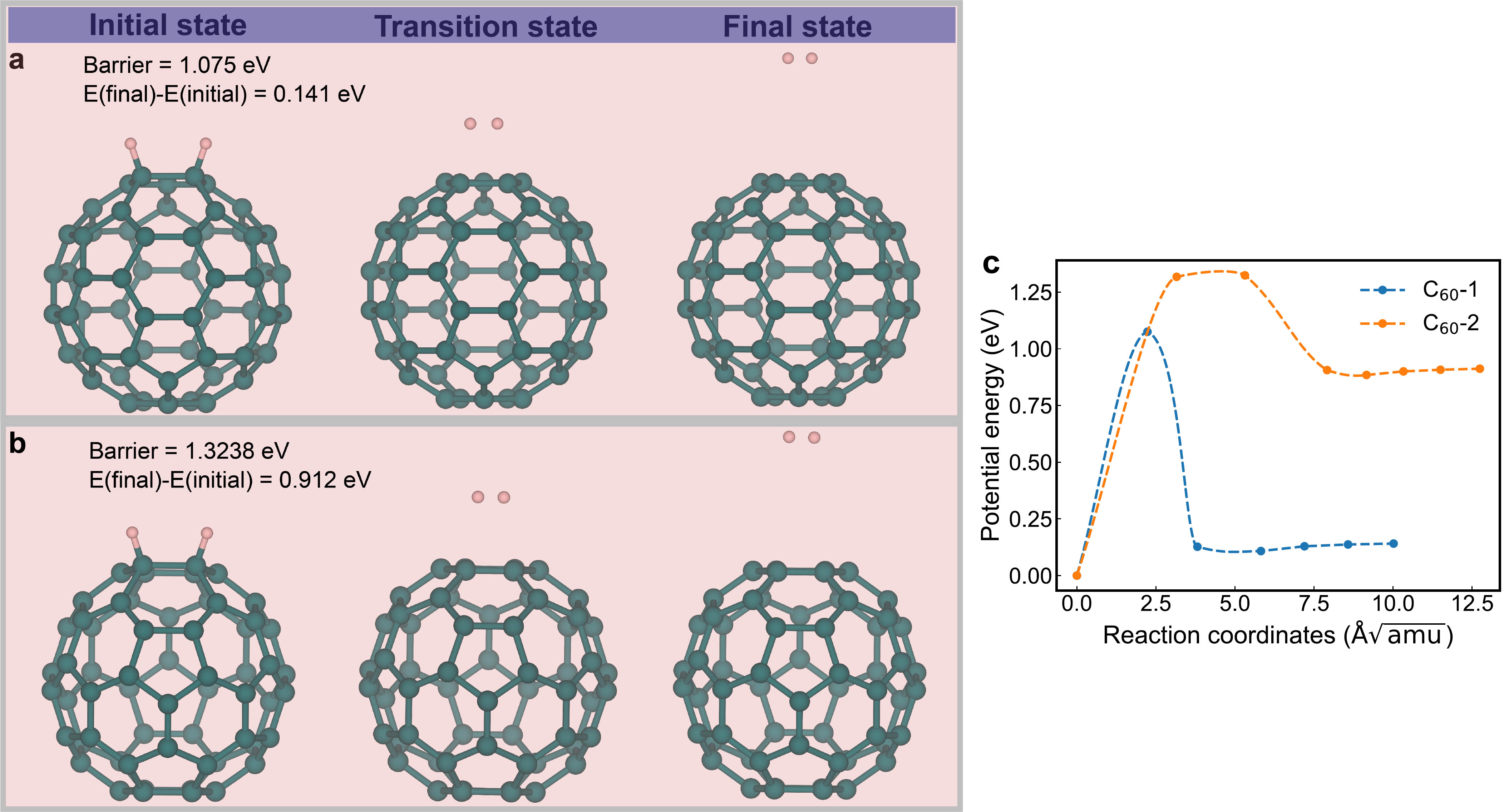}
    \caption{
    Same as Supplementary Fig.~\ref{fig7} for fullerene (C$_{60}$).
    }
    \label{fig10}
\end{figure}


\clearpage
\section{Tests with a hybrid exchange correlation functional}
\label{sec5}

In this section we examine the impact of self interaction errors on our results by comparing with a hybrid exchange correlation functional, namely the HSE06 functional \cite{krukau_influence_2006} (with D3 dispersion corrections \cite{grimme_consistent_2010}).
First we tested the impact on the potential energy barriers.
One can see from Supplementary Table~\ref{tab5} that the barriers computed with the HSE06-D3 are slightly higher than the barriers computed with PBE-D3 for the ortho pathway and the para pathway.
Little change is observed for the meta pathway, therefore we focus our discussion on the other two pathways in this section.
We do not see any qualitative change on the potential energy barriers when switching to a hybrid functional.
\begin{table}[!ht]
\caption{
Potential energy barriers for H$_2$ formation processes calculated using PBE-D3 and HSE06-D3 functional.
The HSE06-D3 results are computed on the geometries optimised with the PBE-D3 functional.\\
}
\begin{tabular}{l|cc}
\hline\hline
\multicolumn{1}{c|}{\multirow{2}{*}{Process}} & \multicolumn{2}{c}{Potential energy barrier (eV)} \\ \cline{2-3} 
\multicolumn{1}{c|}{} & \multicolumn{1}{c}{PBE-D3} & \multicolumn{1}{c}{HSE06-D3} \\ 
\hline
Ortho & 2.389 & 2.776 \\
Meta & 0.756 & 0.725 \\
Para & 1.234 & 1.460 \\ 
\hline\hline
\end{tabular}
\label{tab5}
\end{table}



Then we examine the impact of using a hybrid functional on the instanton rate.
We employed the correction method described in ref.~\cite{Kastner_2018} to estimate the instanton rate with the HSE06-D3 functional.
Specifically, we computed the potential energy using HSE06-D3 functional along the instanton trajectory optimised with the PBE-D3 functional (see Supplementary Fig.~\ref{fig11}).
This allows us to estimate the difference in the $S=W+\beta\hbar E_\mathrm{I}$ action along the instanton trajectory between the two functionals ($\Delta S$).
The factor $\mathrm{e}^{-\Delta S/\hbar}$ is the correction factor to the instanton rate.
For H$_2$ formation on the ortho configuration at 150 K, the instanton rate 
changes from 3.25$\times$10$^{-12}$~s$^{-1}$ to 1.59$\times$10$^{-13}$~s$^{-1}$ switching from PBE-D3 to HSE06-D3. 
For the para pathway, the instanton rate at 100 K changes from 6.82$\times$10$^{-10}$ s$^{-1}$ to 9.28$\times$10$^{-13}$ s$^{-1}$.
These changes do not have any qualitative impact on our findings, and quantitatively, they bring the rates of the ortho path even closer to the rates of the para path, further strengthening the finding that the ortho path is important in the quantum scenario.
We also computed the tunnelling factors with HSE06-D3. 
For H$_2$ formation on the ortho configuration, the tunnelling factor calculated using the PBE-D3 functional is 1.33$\times$10$^{44}$, while the value calculated using the HSE06-D3 functional increases to 6.51$\times$10$^{55}$. 
The PBE-D3 and HSE06-D3 tunnelling factors for H$_2$ formation on the para configuration are 1.16$\times$10$^{28}$ and 3.96$\times$10$^{36}$, respectively. 
%
\begin{figure}[!ht]
    \centering
    \includegraphics[scale=0.085]{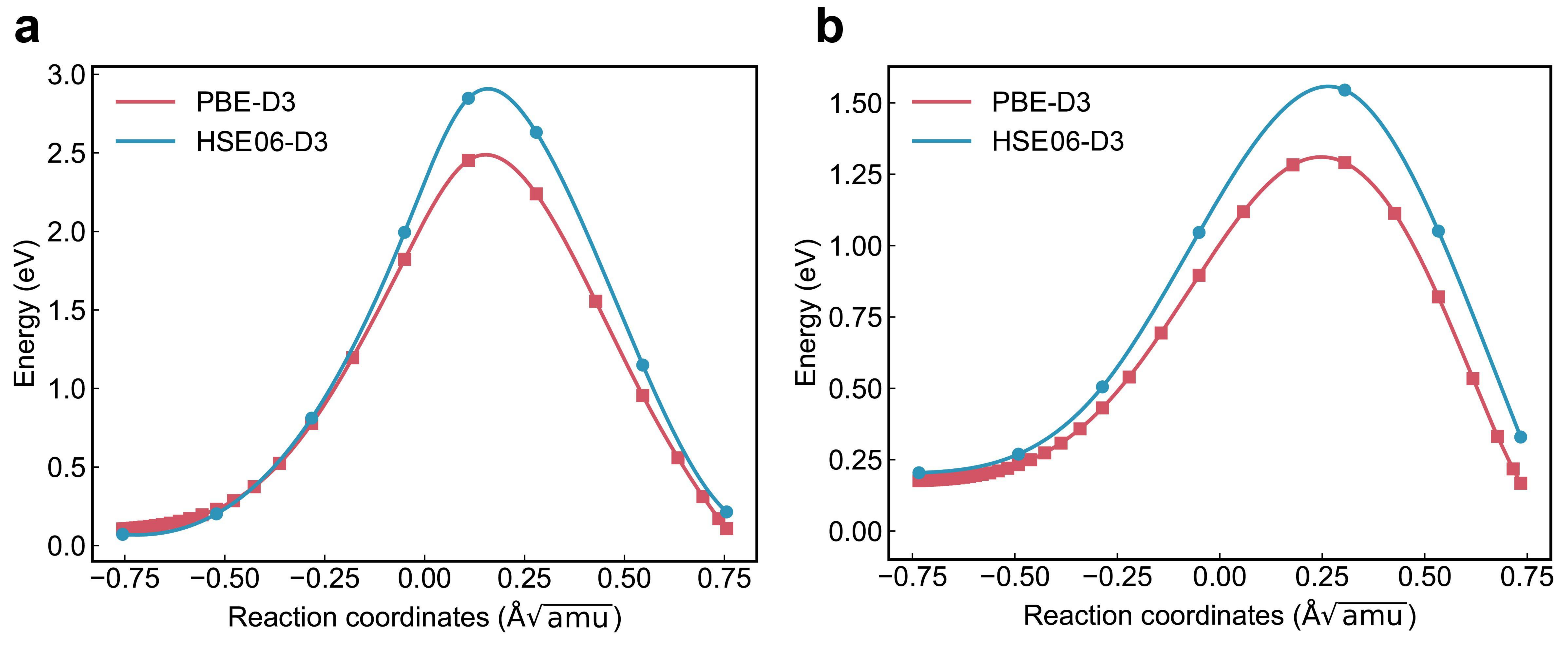}
    \caption{
    Potential energy profile along the instanton trajectories
    for H$_2$ formation on (a) the ortho configuration at 150 K, and (b) the para configuration at 100 K\@. 
    The HSE06-D3 functional results are computed using the instanton geometries optimised with the PBE-D3 functional.
    }
    \label{fig11}
\end{figure}

\clearpage

\bibliographystyle{achemso}
\bibliography{ref-si.bib}